\documentclass[12pt,psf,epsf]{article}

\usepackage[centertags]{amsmath}
\usepackage{graphicx}
\usepackage{epsfig}
\usepackage{ulem}
\usepackage[english]{babel}
\usepackage{array}
\usepackage{amsthm}
\usepackage{latexsym}
\usepackage[mathcal]{euscript}
\pdfoutput=1
\usepackage{epsfig}
\usepackage{color}
\usepackage{jheppubm}
\usepackage{hyperref}
\usepackage{bm}
\usepackage{circledsteps}
\usepackage[dvipsnames]{xcolor}
\usepackage{float}
\usepackage{tikz}
\usetikzlibrary{decorations.pathreplacing}

\definecolor{CustomGreen}{rgb}{0, 0.6, 0.5}

\newcommand{\be}{\begin{equation}}
\newcommand{\ee}{\end{equation}}
\newcommand{\bea}{\begin{eqnarray}}
\newcommand{\eea}{\end{eqnarray}}
\newcommand{\nn}{\nonumber}
\definecolor{darkblue}{rgb}{0., 0, 1}

\title{Lifshitz-like  black branes in arbitrary dimensions
and the third law of thermodynamics
}
\author{Irina Ya. Aref'eva$^{a,b}$, Anastasia A. Golubtsova$^c$ and Valeriya D. Nerovnova$^d$}
\affiliation{$^a$Steklov Mathematical Institute, Russian Academy of Sciences, Gubkina str. 8, 119333, Moscow, Russia\\
$^b$ Moscow State University, Physical Department, Moscow, Russia\\
$^c$ Bogoliubov Laboratory of Theoretical Physics, JINR,
Joliot-Curie str. 6, 141980, Dubna, Russia \\
$^d$Moscow Institute of Physics and Technology, Institutsky lane, no. 9, 141701, Dolgoprudny, Moscow region, Russia \\
}

\emailAdd{arefeva@mi-ras.ru}
\emailAdd{golubtsova@theor.jinr.ru}
\emailAdd{nerovnova.vd@phystech.edu}
\abstract{
In this paper, we systematically construct both isotropic and anisotropic black brane solutions in arbitrary spacetime dimensions $D$, focusing particularly on Lifshitz-like asymptotics.
Two distinct holographic models are considered. The first model involves a scalar field with a potential coupled to two Maxwell fields, allowing for both electric and magnetic charges. The second model includes a scalar field, a Maxwell field, and a three-form field strength of a Kalb-Ramond field. For each model, we derive exact solutions for the metric, scalar field, gauge fields, and coupling functions, incorporating anisotropic scaling exponents and general warp factors, including Gaussian forms.
These results generalize previously known five-dimensional anisotropic black brane solutions to arbitrary dimensions. We show that the third law of thermodynamics, which requires entropy to vanish as temperature approaches zero, is satisfied for a certain range of parameters in both models. However, for specific warp factors or coupling constants, the entropy-temperature relation exhibits non-monotonic or multi-valued behavior, suggesting the possibility of phase transitions and a violation of the third law.}

\begin{document}
\maketitle

\newpage
\section{Introduction}

There are striking parallels between the mechanics of black holes and the laws of thermodynamics \cite{Bardeen:1973gs}. The zeroth law of thermodynamics, according to which the temperature $T$ is constant throughout a system in thermal equilibrium, is analogous to the fact that the surface gravity $\kappa$ is constant over the event horizon of a stationary black hole. This suggests that surface gravity plays the role of temperature.
The first law, according to which the change in internal energy is equal to the heat added plus the work done, is analogous to the relation connecting the change in mass with a 'heat' term plus rotational and electric work.
The second law, which asserts that entropy never decreases $\Delta S \ge 0$, has an analog in the area theorem \cite{Hawking:1971vc}, which states that the horizon area never decreases.
These analogies led Bekenstein \cite{Bekenstein:1973ur} to argue that black holes are genuine thermodynamic systems and that their entropy is proportional to their horizon area.

The third law of thermodynamics, in its Planck formulation, states that entropy $S \to 0$ as $T \to 0$. This formulation does not hold for the Schwarzschild  black hole\footnote{There are at least two formulations of the third law in thermodynamics: the Planck formulation and the Nernst formulation. The Nernst formulation states that it is impossible to reach absolute zero temperature in a finite number of steps. In black hole mechanics, Bardeen, Carter, and Hawking  formulated \cite{Bardeen:1973gs} the third law as: "No finite sequence of physical processes can reduce the surface gravity $\kappa$ to zero." This is a statement of unattainability and is the analogue of the Nernst formulation, not the Planck one.
Note also the D'Hoker and Kraus \cite{DHoker:2009ixq} formulation of third law that that bases on the stability of the extremal (zero-temperature) horizon.}.
This was recognized early on, and alternative formulations of the third law for black holes have since been proposed \cite{Israel:1986gqz,Wald:1997qp}.

In \cite{DHoker:2009ixq}, the third law is considered within the AdS/CFT correspondence framework.
 It is argued that models in which the third law is violated classically, with $S>0$ at $T=0$, are unstable. General perturbations drive the system out of this state \footnote{ D'Hoker and Kraus demonstrated that a specific class of extremal black branes with $S > 0$ at $T = 0$ are unstable under certain generic perturbations (specifically, the introduction of a magnetic field).
However, the statement should not be generalized to all theories or all extremal black holes with nonzero entropy without these qualifiers.}.
Thus, the third law is reformulated as a condition for the existence of a stable ground state in the dual field theory. This formulation is consistent with the idea that the true ground state of a physical system must be stable, and that a stable state at zero temperature will possess the thermodynamic properties required by the third law.

 To compute black hole or black brane entropy from    the statistical mechanics  of ordinary matter \cite{Landau-2013}, it is natural to consider only black holes or branes that satisfy the usual third law. Such an approach may also help illuminate and potentially resolve the information paradox.
Various approaches have been pursued to describe black holes using ensembles of dynamical systems, including shells, D-branes, matrices, Bose gases, and others. The first of these was a controlled calculation of the black brane entropy, performed using methods based on the D-brane/string duality \cite{Strominger:1996sh}. 
 This calculation was followed by many similar computations of entropy for large classes of extremal and near-extremal black holes, and the results consistently agreed with the Bekenstein–Hawking formula.
However, for the Schwarzschild black hole — the furthest-from-extremal black hole — the relationship between microstates and macrostates remains unexplored. 
It can be speculated that this is due to the impossibility of fulfilling the third law in Planck form: entropy explodes at $T\to 0$  for a Schwarzschild black hole. In particular, in  \cite{Arefeva:2023kpu,Arefeva:2023kwm} the thermodynamical behaviour of various  black hole solutions were given in terms of Bose gas models. In this approach, the violation of the third law in the  Schwarzschild black hole thermodynamics has been  explained by a negative dimension of the space\footnote{Physical quantities such as free energy and entropy in negative dimensions have been interpreted in \cite{Arefeva:2023kpu,Arefeva:2023kwm} through analytic continuation, mirroring the use of non-integer dimensions in the dimensional regularization of 't Hooft and Veltman in QFT, and in Wilson's approach to calculating critical exponents in phase transitions.} in a dual Bose gas model that recovers the  Schwarzschild black hole thermodynamics. 

In contrast, for certain black brane solutions — such as Poincare AdS black branes,
Lifshitz black branes, and anisotropic Lifshitz-like black branes — the third law is preserved, with the entropy vanishing as the temperature approaches zero \cite{Arefeva:2024wqb}. For these models, a duality  between these black branes and Bose gas thermodynamics has been found; specifically,
the duality between Lifshitz branes and Bose gases  of quasi-particles with the energy depending on the  Lifshitz parameter $\alpha$.

The goal of this paper is to find a more general class of $D$-dimensional black brane solutions  that satisfy the classical third law. 
The main characteristics of these solutions is their  anisotropic asymptotics. First, we find  $D$-dimensional versions of the 5-dimensional black branes discussed in \cite{Arefeva:2016phb,Arefeva:2018hyo}, see also \cite{Pang:2009ad,Goldstein:2009cv,Azeyanagi:2009pr,Arefeva:2014vjl,Taylor:2015glc,Kubiznak:2016qmn,Giataganas:2017koz,Arefeva:2023jjh,Arefeva:2024mtl,Giataganas:2025ing}. The model under consideration  generalizes the model from \cite{Arefeva:2016phb}  with a magnetic and scalar fields adding a non-trivial scalar potential and an electric ansatz for the Maxwell fields. Then, we consider anisotropic black brane solutions to a $D$-dimensional gravity model including Maxwell, Kalb-Ramond fields and a scalar field with its potential. For the case of vanishing scalar potential, the model turns to be similar to the action describing $p$-branes \cite{Duff:1993ye,Lu:1995sh,Arefeva:1996sjv}. The metrics arising from these models have a scaling exponent analogous to the one found in Lifshitz solutions \cite{Pang:2009ad,Taylor:2015glc}. However, the geometries of the Lifshitz black branes and the black holes from \cite{Arefeva:2016phb} are different: the latter assume a boost-invariance, while for the Lifshitz solutions  Lorentz invariance is violated.  We call the black hole solutions of our interest Lifshitz-like backgrounds following to the work \cite{Azeyanagi:2009pr}.
 Moreover, for the model with two- and three-form fields, we introduce an additional anisotropic parameter $c_B$ for one of the spacial component of the metric.

We show that the third law of thermodynamics, which requires the entropy to vanish as the temperature approaches zero, is satisfied for a certain range of parameters in both models. However, for specific warp factors or coupling constants, the entropy–temperature relation exhibits non-monotonic or multi-valued behavior, suggesting the possibility of phase transitions and a violation of the third law.
More precisely, in the first model (Sect.\ref{sec: section 2}), the pure magnetic black brane solutions with a warp factor $b=1$ obey the third law. In the same model with a nonzero electric field, the third law holds only for certain relations between $D$, the anisotropic parameter $\nu$, and the dilaton coupling constants $k$. Including the Gauss warp factor for the magnetic solutions of the first model leads to a violation of the third law. For the second model (Sect.\ref{sec: section 3}), for the black brane solutions with the Gauss warped factor and both non-zero two- and three- form fields, the third law hold only for a special relation between parameters ($D$, $c$, $c_B$ and $\nu$), the dependence of the entropy on the temperature is generally non-monotonic.

The paper is organized as follows.
In Section \hyperref[sec: section 2]{2} we consider a $D$-dimensional gravity model with two Maxwell fields and a scalar field with a potential. We derive the equations of motion and construct  black brane solutions  with Lifshitz-like anisotropy and a warped-factor $b$ for an arbitrary dimension $D$ in Section \hyperref[subsec: section 2.3]{2.3}. We discuss thermodynamics for pure magnetic and electro-magnetic black branes with $b=1$ and $b=e^{c z^2}$ in Sections \hyperref[subsec: section 2.5]{2.5} and \hyperref[subsec: section 2.6]{2.6}, correspondingly. In Section \hyperref[sec: section 3]{3} we discuss a $D$-dimensional gravity model with two- and three-form field strengths, a scalar field and  a potential.  We construct   $D$-dimensional black brane solutions with $b=e^{-cz^2}$   in Section \hyperref[subsec: section 3.3]{3.3} and its thermodynamics for a certain set of parameters in Section \hyperref[subsec: section 3.5]{3.5}. Then, we discuss particular solutions for $D=5$ in Section \hyperref[subsec: section 3.4]{3.4}.   In Appendix \hyperref[sec: appendix A]{A}, we leave technical details to derive equations of motion. In Appendix \hyperref[sec: appendix B]{B}, we present components of the stress-energy tensors for the first and the second models.
\newpage

\section{Holographic model with Maxwell and scalar fields in arbitrary dimensions}\label{sec: section 2}

\subsection{The setup}

The first model of our interest is a $D$-dimensional generalization of a 5-dimensional Einstein-dilaton-Maxwell theory from \cite{Arefeva:2018hyo}.
The action of the model is given by
\be
      S = \int d^Dx\, \sqrt{-g} \left(
    R - \cfrac{f_1(\phi)}{4} \ F_{(1)}^2 -  \cfrac{f_2(\phi)}{4} \
    F_{(2)}^2 - \cfrac{1}{2} \ \partial_{\mu} \phi \partial^{\mu} \phi - V(\phi) \right), \label{action: el-mag fields model}
\ee
where $R$ is the Ricci scalar,  $g$ is the determinant of the metric, $\phi$ is a scalar field, $V(\phi)$ is its potential, $F_{(i)}$ with $i=1,2$ are Maxwell fields, such that $F_{(1)}$  has an electric ansatz 
\be
F_{\mu\nu} = \partial_\mu A_\nu - \partial_\nu A_\mu, 
\ee
and $F_{(2)}$ is magnetic, for both of them we have
\be
F^{2}_{(i)} =F_{(i)\mu\nu}F_{(i)\rho\sigma}g^{\mu\rho}g^{\nu\sigma},
\ee
$f_i(\phi)$, $i=1,2$, are kinetic functions associated with the corresponding Maxwell fields.

The generic form of the Einstein equations 
read
\be\label{EinEq}
 R_{\mu\nu} - \cfrac{1}{2}g_{\mu\nu}R = T_{\mu\nu},
\ee
where $R_{\mu\nu}$, $R$ are calculated on the metric $g_{\mu\nu}$ and $T_{\mu\nu}$ is the stress-energy tensor is  given by
\be
T_{\mu\nu}=\frac{1}{2}\left(\partial_{\mu}\phi\partial_{\nu}\phi-\frac{1}{2}g_{\mu\nu}\partial_{\rho}\phi\partial^{\rho}\phi -g_{\mu\nu}V(\phi)\right)+\frac{f_i(\phi)}{2}\left(-\frac{1}{4}g_{\mu\nu}F_{(i)}^2+F_{\mu\rho(i)}F^{\rho}_{(i)\nu}\right).
\ee

The scalar field equation can be represented in the following form
\be\label{dilaton EOM: general}
\Box \phi= \cfrac{1}{4}\cfrac{\partial f_1}{\partial\phi}F^2_{(1)} + \cfrac{1}{4}\cfrac{\partial f_2}{\partial\phi}F^2_{(2)} + \cfrac{\partial V}{\partial\phi}, \quad \Box=\frac{1}{\sqrt{|g|}}\partial_{\mu}\left(g^{\mu\nu}\sqrt{|g|}\partial_{\nu}\right).
\ee
The equations for the Maxwell fields read
\be\label{field EOM: general}
\partial_\nu(\sqrt{-g}\;f_i(\phi)F_{(i)}^{\mu\nu}) = 0\,,
\ee
where $i = 1,2.$

We consider the ansatz of the black brane metric, which  depends only on  the radial coordinate $z$ and is taken in the following form:
\be
    ds^2 = \frac{L^2 \, b(z)}{z^2} \left( - \ g(z) dt^2 + dx_1^2 + ... +dx_d^2 + 
    z^{2-\frac{2}{\nu}} \left( dy_1^2 + dy_2^2 \right) +
    \cfrac{dz^2}{g(z)} \right), \label{D-metric: el-mag fields}    
\ee
where $g(z)$ is the blackening function, which vanishes on the horizon and goes to $1$ on the boundary of the spacetime
\be
g(0) = 1,\quad g(z_h) = 0,\label{boundary conditions}
\ee
with the black brane horizon $z_h$. In \eqref{D-metric: el-mag fields}
$b(z)$ is a warp factor, $\nu$ is a parameter of anisotropy.  The isotropic ansatz with $\nu = 1$ and $b = 1$ yields an asymptotically AdS spacetime. Interestingly, for $b = 1$ and arbitrary $\nu$, the metric \eqref{D-metric: el-mag fields} asymptotes to $AdS_{D-2} \times M_2$ near the boundary as $z \to 0$.

The magnetic forms are located on the  $y_1$ and $y_2$ directions
\be\label{F2ansatz}
F_{(2)} = q \,dy_1\wedge dy_2,
\ee
where  $q$ is a constant
and we assume that the vector-potential $A_t$ and the scalar field $\phi$   have dependence only on the radial coordinate $z$
\be\label{scale}
\phi = \phi(z),\quad A_\mu = \delta^0_\mu A(z).
\ee

\subsection{The equations of motion}

In this subsection we discuss the EOM on our ansatz of the metric and the fields \eqref{D-metric: el-mag fields}-\eqref{scale}.
Doing some algebra, we are brought to the following combinations of Einstein equations
\bea
&&b''- \cfrac{3(b')^2}{2b} + \cfrac{2b'}{z} + \cfrac{4b}{z^2\nu^2(D-2)}(1 - \nu) + \cfrac{b}{D-2}(\phi')^2 = 0\,,\quad\quad  
\label{dilaton equation: el-mag-fields model}\\
&&g'' + g'\left(\cfrac{b'}{2b}(D-2) - \cfrac{2}{z\nu} - \frac{(D-4)}{z}\right) - \cfrac{z^2f_1(A_t')^2}{b} = 0\,, \quad\label{g equation: el-mag-fields model}\\
&&\left(1 - \cfrac{1}{\nu}\right)\left(2g' + g\left(\cfrac{b'}{b}(D-2) - \cfrac{4}{z\nu} - \cfrac{2(D-3)}{z}\right)\right) + \cfrac{q^2z^{-1+\frac{4}{\nu}}f_2}{b} = 0\,, \label{f2 equation: el-mag-fields model}\\
&&\cfrac{b''}{3b}(D-2) + \cfrac{(b')^2}{6b^2}(D-2)(D-4) + \cfrac{g''}{3g} + g'\left(\cfrac{b'(D-2)}{2bg} 
+ \cfrac{\nu(20-6D) - 8}{6z\nu g}\right) \nn \\&&+ \cfrac{2bV}{3z^2g} - {\cfrac{{b'}(D-2)}{{b}z\nu}} - {\cfrac{{b'}(D-2)(2D - 7)}{3{b}z}} 
+ \cfrac{2(2 + 3\nu(D-3) + \nu^2(D-3)^2)}{3z^2\nu^2} = 0.\qquad\qquad\label{V equation: el-mag-fields model}
\eea

We present non-zero components of the $D$-dimensional Einstein tensor on the ansatz \eqref{D-metric: el-mag fields}-\eqref{scale} in Appendix \hyperref[sec: app.A]{A}.  

The field equations for the scalar field \eqref{dilaton EOM: general} and gauge potential \eqref{field EOM: general}  take the form:
\bea\label{phi}
  &&  \phi'' + \phi'\left(\cfrac{g' }{g} + \cfrac{b'}{b}\left(\cfrac{D}{2} - 1\right) + \cfrac{4 - D}{z}-\cfrac{2}{\nu z}\right) + \cfrac{\partial f_1}{\partial\phi}\cfrac{z^2(A_t')^2}{2bg} - \cfrac{\partial f_2}{\partial\phi}\cfrac{z^{-2 + \frac{4}{\nu}}q^2}{2bg} {- \cfrac{b}{z^2g}\cfrac{\partial V}{\partial \phi} } = 0,\qquad\quad \\
    && A_t'' + A_t'\left(\cfrac{b'}{b}\left(\cfrac{D}{2}-2\right)+\cfrac{f_1'}{f_1} - \cfrac{2 - \nu(6 -D)}{\nu z}  \right) = 0.\label{At equation: el-mag-fields model}
\eea

The equation for the field strength $F_{(2)}$ defined by \eqref{F2ansatz}  is given by
\be
\partial_\mu(\sqrt{-g}f_2(\phi(z))F^{\mu\nu}_{(2)})\;\equiv\;0.
\ee

\subsection{Generic black brane solutions}\label{subsec: section 2.3}
In this subsection we will construct  exact black brane solutions to eqs.\eqref{dilaton equation: el-mag-fields model}-\eqref{At equation: el-mag-fields model}
with an arbitrary function $b(z)$.

First, we focus on the case $A_t = 0$. The  solution for the blackening function with an arbitrary $b(z)$ is obtained from \eqref{g equation: el-mag-fields model} such that the boundary conditions \eqref{boundary conditions} are satisfied, i.e.:
\be
g_0(z) = 1 - \cfrac{ \int\limits_0^z \cfrac{t^{D+\frac{2}{\nu} - 4 }}{b^{ \frac{D}{2}-1  }  } dt}{\int\limits_0^{z_h} \cfrac{ t^{D+\frac{2}{\nu} - 4 } }{ b^{ \frac{D}{2}-1 }  } dt}\label{g solution: magnetic black brane el-mag model}.
\ee
Index ``0'' in the formula above corresponds to zero value of the chemical potential $\mu$.\\

The scalar field  can be found from \eqref{dilaton equation: el-mag-fields model} with the reality condition for the scalar field solution
\be
\cfrac{4}{z^2\nu^2}(\nu - 1) - \cfrac{b'}{b}\cfrac{2(D-2)}{z} + \cfrac{3}{2}(D-2)\left(\cfrac{b'}{b}\right)^2 - \cfrac{b''}{b}(D-2) \geqslant 0,
\ee
such that it reads
\be
\phi = \pm\int \sqrt{ \cfrac{4}{z^2\nu^2}(\nu - 1) - \cfrac{b'}{b}\cfrac{2(D-2)}{z} + \cfrac{3}{2}(D-2)\left(\cfrac{b'}{b}\right)^2 - \cfrac{b''}{b}(D-2) }\,dz + \phi_0,
\ee
where $\phi_0$ is a constant of integration that depends on the boundary conditions for $\phi$.

The solution for the coupling function $f_2$ is derived in the following form:

\be
f_2= \cfrac{ z^{-\frac{4}{\nu}}(\nu -1) }{q^2\nu }\left(1 - \cfrac{ \int\limits_0^z \cfrac{t^{D+\frac{2}{\nu} -4} }{b^{\frac{D}{2} -1}  } dt}{\int\limits_0^{z_h} \cfrac{t^{D+\frac{2}{\nu} -4} }{ b^{ \frac{D}{2}-1 }  } dt}\right)\left[2b\left(\cfrac{2}{\nu} + D-3\right) -\right. \left.2z\cfrac{ \cfrac{ z^{D+\frac{2}{\nu} -4} }{b^{ \frac{D}{2}-2 }  }    }{\int\limits_{z_h}^z \cfrac{t^{D+\frac{2}{\nu} -4}}{b^{ \frac{D}{2}-1 } } dt} - b'z(D-2)\right].
\ee
Finally, the scalar potential is given by
\bea
V(z) &=& \cfrac{g}{b}\Big[\cfrac{1}{\nu}\Bigl( 9 - \cfrac{2}{\nu} - 3D\Bigr)-(D-3)^2 + \cfrac{b'}{b}z\Bigl(7 - \cfrac{11D}{2}+D^2 + \cfrac{3}{2\nu}(D-2)\Bigr) \\
&-&  \cfrac{1}{4}\Bigl(\cfrac{b'}{b}z\Bigr)^2(D-4)(D-2)\Big] + \cfrac{g'}{b}z\left[ D-3+\cfrac{1}{\nu} - \cfrac{b'}{2b}z(D-2) \right] - g z^2\cfrac{b''}{2b^2}(D - 2).\nn
\eea

Now we consider the case of a non-zero electric field.
If $A_t\neq 0$ the equation for the blackening function \eqref{g equation: el-mag-fields model} appears to be inhomogeneous and should be solved along with the EOM for the vector potential $A_t(z)$. Having boundary conditions for $A_t$
\be
A_t(0) = \mu, \quad A_t(z_h) = 0, \label{boundary conditions for A}
\ee
where $\mu$ is a chemical potential, from \eqref{At equation: el-mag-fields model}, we have:
\be
A_t(z) = A_0\int_{z_h}^z\cfrac{t^{\frac{2}{\nu} +D-6}}{f_1(t)\,b^{\frac{D}{2}-2} }dt, \quad A_0 = \cfrac{\mu}{\int_{z_h}^0 \frac{t^{\frac{2}{\nu} + D - 6 } }{f_1(t)\,b^{\frac{D}{2}- 2}} dt}.\label{At solution: el-mag-fields model}
\ee
The blackening function is obtained from the equation \eqref{g equation: el-mag-fields model} and given by the formula:
\bea
&&g(z) 
 = G_0 g_0(z) + g_1(z)\,,\label{electric solution}
\eea
where $g_0(z)$ is the solution of the homogeneous equation \eqref{g solution: magnetic black brane el-mag model} (without the electric field), $G_0$ is a constant, which depends on the location of the horizon $z_h$, so
\be
G_0 = 1-A^2_0\int\limits_0^{z_h} \cfrac{t^{ D-6+\frac{2}{\nu} }  }{f_1(t)b^{\frac{D}{2}-2}}\left[\int\limits_t^\infty h(u)du - \int\limits_0^\infty h(\Tilde{u})d\Tilde{u}\right]dt\,,\label{electric solution scaling constant}
\ee
where $A_0$ is given  by \eqref{At solution: el-mag-fields model}. The function $g_1(z)$ in \eqref{electric solution} is defined by
\be\label{gonegen}
g_1(z) = A_0^2\int\limits_{z_h}^z \cfrac{t^{ D-6+\frac{2}{\nu} } }{f_1(t) b^{\frac{D}{2}-2}}\left[\int\limits_t^\infty h(u)du - \int\limits_z^\infty h(\Tilde{u})d\Tilde{u}\right]dt,
\ee
where $h(z) = \cfrac{z^{D-4+\frac{2}{\nu}}  }{b^{\frac{D}{2}-1}(z)  }$.

\subsection{Special black brane solutions with $b(z) = 1$}

\subsubsection{Magnetic black brane}

Let us focus on the case of zero electric field $F_{(1)} = 0$ and the simplest choice $b(z) = 1$. Then, from eqs. \eqref{dilaton equation: el-mag-fields model}-\eqref{V equation: el-mag-fields model} and  \eqref{boundary conditions} we find the following magnetic black brane solution:
\be \label{magneticb1}
ds^2 = \frac{L^2 \, }{z^2} \left( - \ g(z) dt^2 + dx_1^2 + ... +dx_d^2 + 
    z^{2-\frac{2}{\nu}} \left( dy_1^2 + dy_2^2 \right) +
    \cfrac{dz^2}{g(z)} \right),   
\ee
with the blackening function
\be
g(z) = 1 - \left(\cfrac{z}{z_h}\right)^{D-3+\frac{2}{\nu}},\label{blackening func solution b=1: el-mag-fields model}
\ee
and the scalar field given by
\be\label{phisolmagbr5}
\phi = \pm \cfrac{2}{\nu}\sqrt{\nu - 1}\log {z} + c_1,
\ee
where $c_1$ is a constant that depends on boundary conditions.

The coupling function $f_2$ \eqref{action: el-mag fields model} reads
\be
f_2(z) = \cfrac{2(\nu - 1)}{q^2\nu^2}(2 + (D-3)\nu)z^{-\frac{4}{\nu}}\,,\label{f2 magnetic solution: el-mag-model}
\ee
and the magnetic field $F_{(2)}$ is defined by \eqref{F2ansatz}.

The scalar potential turns to be constant and reads
\be\label{Vmagnetic}
V = -\cfrac{(1 + \nu(D-3))(2 + \nu(D-3))}{\nu^2}.
\ee
 The black brane  \eqref{magneticb1}-\eqref{Vmagnetic} is a generalization of the $5d$ black brane solution from in \cite{Arefeva:2016phb} to the case of an arbitrary dimension. To match with the solution from   \cite{Arefeva:2016phb} we choose the coupling function $f_2$ as $f_2(\phi) = e^{\lambda\phi}$ with   a dilatonic coupling constant $\lambda$. Setting $\lambda$ as
\be
\lambda = \pm \cfrac{2}{ \sqrt{\nu - 1} },
\ee
we read off the constraint for the parameter $q$:
\be
q^2 = \cfrac{2(\nu - 1)(2 + (D-3)\nu)}{\Tilde{c_1}\nu^2},
\ee
where $\Tilde{c}_1$ is a constant that depends on boundary conditions for $\phi: \; \Tilde{c}_1 = e^{\lambda c_1}$.

We can see that for $\nu=1$  the scalar field $\phi$  and $f_2$ vanish while the constant scalar potential $V$ becomes a cosmological constant $V = -\Lambda$ and the solution turns into an ordinary AdS black brane.

Note that if the blackening function takes  $g=1$ the metric \eqref{magneticb1} is a $d$-dimensional version of the Lifshitz-like metric discussed in the work \cite{Arefeva:2014vjl}.

\subsubsection{Black brane solution with non-zero electric and magnetic charges}

Now we derive the solution for the case where both electric and magnetic charges are nonzero. The metric of the solution is similar to the previous case \eqref{magneticb1}, but has a different form for the blackening function $g$ \eqref{electric solution}.
Taking into account \eqref{electric solution} and \eqref{electric solution scaling constant} with the trivial warp-factor $b(z) = 1$ we find for the blackening function:
\bea
g(z) &=& \left(1 - \left(\cfrac{z}{z_h}\right)^{ \frac{2}{\nu} + D-3 }\right)\left(1 - \cfrac{A_0^2}{\frac{2}{\nu} +D-3}\int_{z_h}^0\cfrac{t^{ \frac{4}{\nu} + 2D-9 }   }{f_1(t)}dt\right) \nn\\
&+& \cfrac{A_0^2}{\frac{2}{\nu}+D - 3}\int_{z_h}^z\cfrac{dt}{f_1(t)}\left[t^{ \frac{4}{\nu}+2D-9 } - z^{\frac{2}{\nu}+D-3 }t^{\frac{2}{\nu}+D-6 }  \right] \nn \\
&=& G_0\left(1 - \left(\cfrac{z}{z_h}\right)^{\frac{2}{\nu}+D-3}\right) + g_1(z)\,,\label{blackening func solution with el. field}
\eea
where the constant $G_0$ is defined by
\be
{G_0} = 1 - \cfrac{A_0^2}{\frac{2}{\nu}+D-3}\int_{z_h}^0\cfrac{t^{ \frac{4}{\nu}+2D-9 } }{f_1(t)}dt\,.
\ee
By virtue EOM \eqref{f2 equation: el-mag-fields model} and doing some algebra we get an explicit form for the  coupling function $f_2$
\bea
f_2(z) = \cfrac{2(1-\nu)}{q^2\nu}z^{-\frac{4}{\nu}}\left[{G_0}\left(\cfrac{2}{\nu}+D-3\right) + A_0^2\int_{z_h}^z\cfrac{t^{ \frac{4}{\nu}+2D-9  }    }{f_1(t)}dt\right]\,.\label{f2 solution At neq 0}
\eea
The scalar field is obtained from \eqref{dilaton equation: el-mag-fields model} and remains the same as in the case of zero electric filed:
\be
\phi(z) 
= -2\lambda\log z + \phi_0, \quad \lambda = \pm \cfrac{\sqrt{\nu -1}}{\nu}. \label{dilaton in el-mag-fields model}
\ee

The scalar potential $V$ can be reconstructed from the solution. 
Thus, the equation for the  potential appears to be the following
\be
9g - 6Dg + D^2g + \cfrac{2g}{\nu^2} - \cfrac{9g}{\nu} + \cfrac{3Dg}{\nu} + V + 5zg' - \cfrac{3}{2}Dzg' - \cfrac{2zg'}{\nu} + \cfrac{1}{2}z^2g'' = 0.
\ee
Combining  the latter  with eq. \eqref{g equation: el-mag-fields model} for the blackening function we have
\be
V= -\cfrac{(1+(D-3)\nu)(2+(D-3)\nu)}{\nu^2}{G_0} - A_0^2\left[\cfrac{ z^{2D+\frac{4}{\nu} - 8}  }{2f_1(z)} + \left(D+\cfrac{1}{\nu}-3\right)\int\limits_{z_h}^z\cfrac{ t^{2D+\frac{4}{\nu}-9} }{f_1(t)}dt  \right].\qquad\quad\label{V solution At neq 0}
\ee

Near the boundary as $g \to 1$  the metric of this  black brane solution  with the electric field has the same asymptotics as in the pure magnetic case discussed earlier.





\subsection{Thermodynamics of black branes with $b(z) = 1$}\label{subsec: section 2.5}

The Hawking temperature of the black brane solutions, which we constructed  in the previous subsection can be found as follows:
\be
T_{H} = \left|\cfrac{g'(z)}{4\pi}\right||_{z\to z_h}.\label{hawking temperature formula}
\ee

The entropy density of the black hole solution is defined through the area of the black brane horizon, i.e.
\be
s =\frac{A}{4V_{D-2}},
\ee
with the area given by
\be\label{entropy formula}
A = \int d\mathrm{x} \sqrt{\gamma},
\ee
where $\gamma$ is an induced metric. 

Particularly, the entropy density of the magnetic black brane solution with the metric \eqref{D-metric: el-mag fields} and the blackening function  \eqref{blackening func solution b=1: el-mag-fields model} reads
\be \label{entropy density: el-mag-fields model}
s(z_h) = \frac{1}{4 z^{D-4+2/\nu}_{h}}.
\ee

\begin{figure}[H]
    \centering
    \includegraphics[scale = 1.]{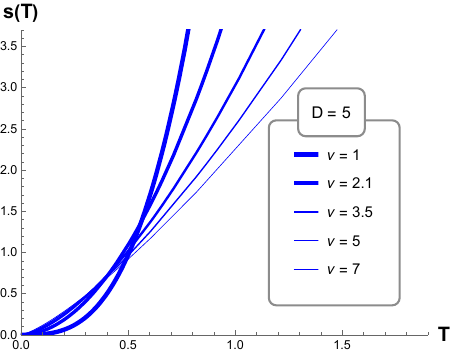}
    \includegraphics[scale = 1.]{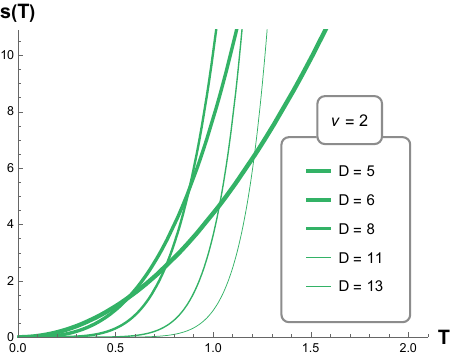}\\
    (A)\hspace{200pt} (B)
    \caption{The entropy density $s$ of the magnetic black brane solutions \eqref{magneticb1}-\eqref{phisolmagbr5} as a function of temperature for different $\nu$ and  $D$. In (A) we fix $D = 5$ and vary   $\nu$, in (B) we fix $\nu = 2$ and vary $D$. The third law is satisfied as the dependence of the entropy density on the temperature is given by a power law.}
    \label{fig: s(T)first}
\end{figure}

Moreover, in this case, the explicit dependence of the entropy density on the Hawking temperature can be derived as follows:
\be\label{entropy on T: el-mag-fields model}
s(T) = \cfrac{1}{4}\left(\cfrac{ \frac{2}{\nu} + D - 3  }{4\pi T}\right)^{4 -D - \frac{2}{\nu}}.
\ee
Since we consider $D\geqslant 5$ and the parameter of anisotropy $\nu \geqslant 1$  for our model, then $4 - D-\frac{2}{\nu} < 0$ and the dependence $s$ on $T $ \eqref{entropy on T: el-mag-fields model} has a power law
\be
s(T)\propto T^{\alpha}, \quad \alpha > 0.
\ee
From the latter, we see that the entropy density $s$ vanishes as $T\to 0$, 
which is consistent with the third law of thermodynamics. We show  the dependence of the entropy density on the temperature \eqref{entropy on T: el-mag-fields model} for different sets of parameters in Figs.~\ref{fig: s(T)first} {\bf (A)} and {\bf (B)}.

As for the case of a non-zero electric field, it can already be shown from the general solution \eqref{electric solution}-\eqref{gonegen}  that the  derivative of $g_1(z)$ tends to zero as $z$ goes to $z_h$ regardless of  the warp-factor $b(z)$ and the coupling function $f_1$, i.e.:
\bea
&&g'_1(z_h) = A_0^2\cfrac{z^{\frac{2}{\nu}+D-6} b^{\frac{D}{2} -2 }}{f_1(z)}\int\limits_z^\infty\cfrac{t^{\frac{2}{\nu}+D-4}  }{b^{\frac{D}{2} - 1} }dt +A^2_0\cfrac{z^{\frac{2}{\nu}+D-4  } }{b^{\frac{D}{2} - 1} }\int\limits_{z_h}^z\cfrac{t^{\frac{2}{\nu}+D-6} b^{\frac{D}{2} -2 } }{f_1(t)}dt -\nn\\
&& -\left.A^2_0\cfrac{z^{\frac{2}{\nu}+D-6} b^{\frac{D}{2}-2 } }{f_1(z)}\int\limits_z^\infty\cfrac{t^{\frac{2}{\nu}+D-4}  }{b^{\frac{D}{2}-1 } }dt\right|_{z = z_h} = 0.
\eea

Therefore,  the Hawking temperature of the black brane solution with the metric \eqref{magneticb1} and the blackening function \eqref{blackening func solution with el. field}  is given by
\bea\label{el-mag model: temperature with el field}
T(z_h) = \cfrac{\frac{2}{\nu}+D-3}{4\pi z_h}\left|1 - \left(\frac{2}{\nu}+D-3\right)^{-1}\left(\cfrac{\mu}{  \int_{z_h}^0\cfrac{ t^{ \frac{2}{\nu}+D-6 }  }{ f_1  }dt }\right)^2\int\limits_{z_h}^0\cfrac{t^{ \frac{4}{\nu}+2D-9 } }{f_1}dt\right|. 
\eea

The entropy density of  the electric black brane with \eqref{magneticb1} and \eqref{blackening func solution with el. field}  has the same form as for the pure magnetic case \eqref{entropy density: el-mag-fields model}  and depends on the  parameter of anisotropy $\nu$.  

Note that the temperature of the black brane solution \eqref{el-mag model: temperature with el field} is related to the choice of the coupling function $f_1$. 
We focus on $f_1 = e^{k (\phi - \phi_0)}$, where $\phi$ and $\phi_0$ are defined by  \eqref{dilaton in el-mag-fields model} and the coupling constant  $k$ is a real parameter.  Therefore, taking into account \eqref{dilaton in el-mag-fields model}  the function $f_1$ takes the form:
\be\label{f1exactmod1}
f_1(z) = z^\kappa,
\ee
where $\kappa = -2k\lambda$ with $\lambda$ from \eqref{dilaton in el-mag-fields model}. Then with \eqref{f1exactmod1} we obtain for the Hawking temperature 
\be\label{TzhElFields1mod}
T(z_h) = \cfrac{\frac{2}{\nu}+D-3}{4\pi z_h} + \cfrac{(D+\frac{2}{\nu}-5-\kappa)^2}{4\pi (2D + \frac{4}{\nu}-8-\kappa) }\mu^2\,z_h^{\kappa+1}.
\ee
Let us introduce a parameter
\be
\sigma = D +\cfrac{2}{\nu} - 3.
\ee
Using \eqref{TzhElFields1mod} and \eqref{entropy density: el-mag-fields model} we can represent the temperature as a function of the entropy in terms of $\sigma$:
\be\label{T(s)ElectricFieldmod1}
T(s) = \cfrac{\sigma}{\pi}\,4^{\frac{2-\sigma}{\sigma - 1}}\, s^{\frac{1}{\sigma - 1}} + \cfrac{\mu^2}{\pi}\,4^{\frac{\kappa + \sigma}{1-\sigma}}\,\cfrac{(\sigma - 2 - \kappa)^2}{(2\sigma - 2 - \kappa)}\,s^{\frac{1+\kappa}{1-\sigma} }.
\ee
With $D\geqslant 5$ and $\nu \geqslant 1$ we have $\sigma > 1$. Therefore, from the expression \eqref{T(s)ElectricFieldmod1} we obtain $\kappa < -1$ to ensure the third law is satisfied. For the given values of $\kappa$, the temperature is defined correctly, as integrals in \eqref{el-mag model: temperature with el field} converge, thus, we get
\be
2k\lambda > 1.
\ee

For certain value of $\kappa$, particularly, for $\kappa = -2$ we can explicitly find $s$ as a function of $T$ and show that the third law is satisfied for any  $z_h$:
\be
s(T) = \cfrac{1}{4}\left[\cfrac{(\frac{2}{\nu}+D-3)}{4\pi T}\left(1+\frac{\mu^2}{2}\right)\right]^{4-D-\frac{2}{\nu}}.
\ee
In this case we recover a  power-law $s(T)\propto T^\alpha$, with $\alpha>0$.

Setting $\kappa  = -2$ fixes the relation between the parameters:
\be
\lambda k = 1,
\ee
that yields $\nu > 1$ for $|k| > 2$. In particular,  we have $k > 2$ for $\lambda > 0$ and $k < -2$ for $\lambda < 0$. Moreover, putting $k = 0$ or $|k| = 2$ corresponds to the isotropic case $\nu = 1$.

\subsection{Magnetic black branes with Gaussian $b=e^{c z^2}$ in $D=5$}\label{subsec: section 2.6}

Now we discuss the model \eqref{action: el-mag fields model} in $D=5$ with the ansatz of the metric \eqref{D-metric: el-mag fields} and  the gaussian warp factor $b(z) = e^{c z^2}$ for the magnetic case. Below we will obtain a solution for the blackening factor $g$ of the magnetic black brane and discuss its thermodynamics.  Note that the scalar field of the solution is defined by eq. \eqref{phi} with $A_t=0$.

We will use the equations of motion \eqref{dilaton equation: el-mag-fields model} - \eqref{V equation: el-mag-fields model} with $b(z) = e^{cz^2}$. The solution for the blackening function can be found from eq. \eqref{g equation: el-mag-fields model} with \eqref{boundary conditions}, so we have
\bea\label{blackfgaussmod}
&&g(z) = \cfrac{\Gamma(\frac{1}{\nu}+1, \frac{3cz^2}{2}) - \Gamma(\frac{1}{\nu}+1, \frac{3cz^2_h}{2})}{\Gamma(\frac{1}{\nu}+1) - \Gamma(\frac{1}{\nu}+1, \frac{3cz^2_h}{2})},
\eea
where $\Gamma(\alpha, x)$ is an upper incomplete gamma-function.

\begin{figure}[H] 
    \centering
    \includegraphics[scale = 1.]{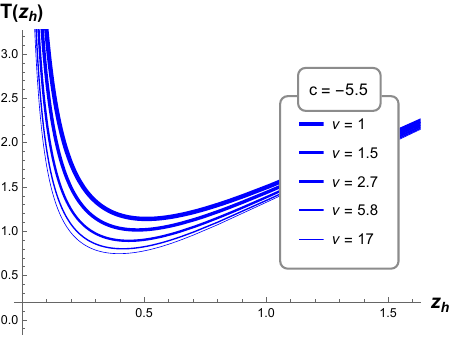}
    \includegraphics[scale = 1.]{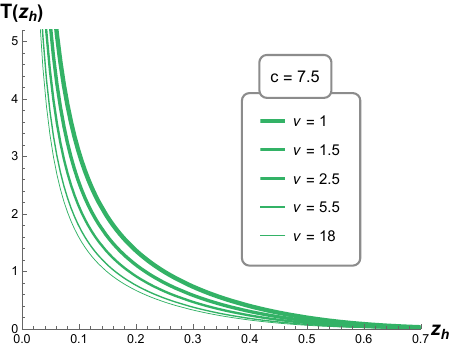}\\
    (A) \hspace{200pt} (B)
    \includegraphics[scale = 1.]{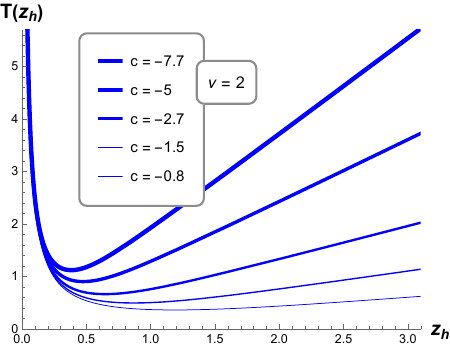}
    \includegraphics[scale = 1.]{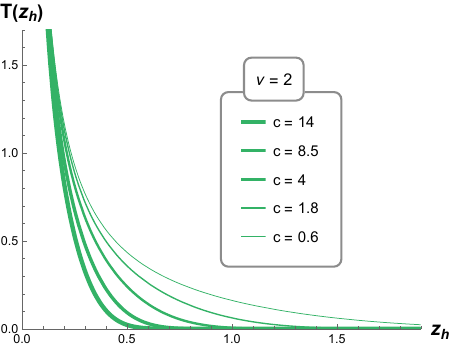}\\
    (C) \hspace{200pt} (D)
    \caption{Hawking temperature as a function of $z_h$ for the 5-dimensional magnetic black brane with the warp factor  $b=e^{cz^2}$ and the blackening function \eqref{blackfgaussmod}. In (A) and (C) we show the Hawking temperature for $c<0$, in (B) and (D) - for $c>0$.}\label{fig: T(zh) el-mag model for D=5}
\end{figure}

The Hawking temperature and the entropy density can be found using \eqref{hawking temperature formula} and \eqref{entropy formula}, correspondingly, i.e.
\bea
&&T(z_h) = \cfrac{2^{-\frac{2}{\nu}} }{4\pi}\left|(3c)^{\frac{1}{\nu}+1}\cfrac{z_h^{ \frac{2}{\nu}+1 } e^{-\frac{3cz^2_h}{2}}  }{\Gamma(\frac{1}{\nu}+1, \cfrac{3cz^2_h}{2}) - \Gamma(\frac{1}{\nu}+1)}\right| ,\label{T: el-mag model D=5}\\
&&s(z_h) = \cfrac{1}{4}e^{\frac{3cz^2_h}{2}}z_h^{-1-\frac{2}{\nu}}.\label{entdengauss}
\eea
We show the temperature \eqref{T: el-mag model D=5} as a function of $z_{h}$ for various values $c$ and $\nu$ in Fig.~\ref{fig: T(zh) el-mag model for D=5} (A)-(D). For (A) and (B) we fix  $c$ and vary $\nu$, for (C) and (D) we  fix $\nu$ and vary $c$. We see that for negative values of $c$ in Fig.~\ref{fig: T(zh) el-mag model for D=5}  (A) and (C) the temperature is non-monotonic and can take the same value for different $z_h$, while for positive $c$ we see that $T$  is a monotonically decreasing function of $z_h$.

\begin{figure}[H]
    \centering
    \includegraphics[scale = 1.0]{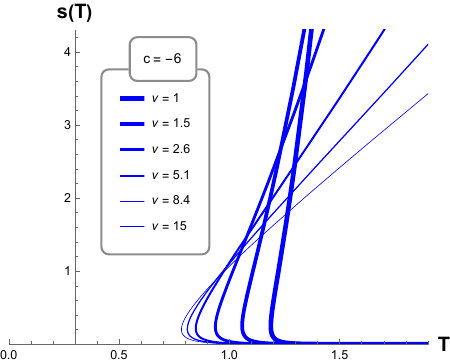}
    \includegraphics[scale = 1.0]{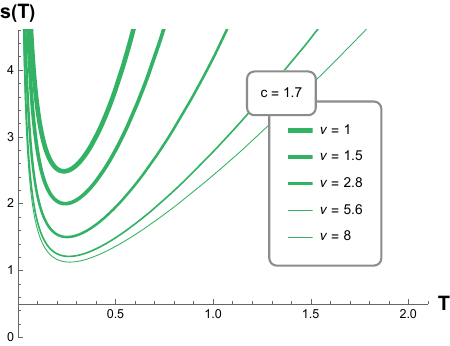}\\
    (A)\hspace{200pt} (B)
    \includegraphics[scale = 1.2]{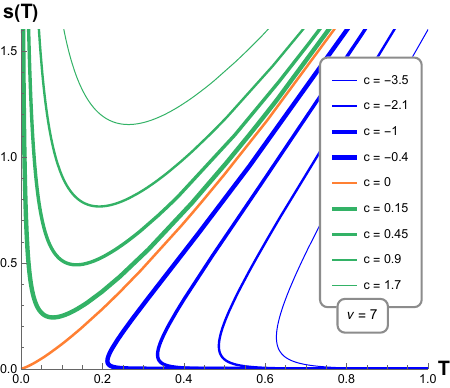}\\
    (C)
    \caption{Entropy density as a function of temperature for the 5-dimensional magnetic black brane with  the  warp factor $b = e^{cz^2}$  and the blackening function \eqref{blackfgaussmod}.  For both Figs. (A) and (B) we vary $\nu$ and fix $c$, negative and positive, correspondingly.  The orange line in Fig. (D) represents $c = 0$ when the third law of thermodynamics is satisfied.}
    \label{fig: s(T) el-mag model D = 5}
\end{figure}

In Fig.~\ref{fig: s(T) el-mag model D = 5} we plot the dependence of the entropy density on the temperature.  For (A) and (B) we fix the parameter $c$ and vary $\nu$.  In Fig.~\ref{fig: s(T) el-mag model D = 5} (C) we vary $c$ with a fixed $\nu$. We see that the entropy depends continuously on the parameter $c$. The value $c = 0$ gives  an almost linear dependence of the entropy density $s$ on the temperature $T$  and separates the curves of $s(T)$ related to negative and positive $c$. Note, that the case with $c=0$ is  the only case which is consistent with the third law of thermodynamics, i.e. $s\to 0$ as $T\to 0$. From Fig.~\ref{fig: s(T) el-mag model D = 5} (C)  we also observe that for negative $c$  a certain value of $T$ corresponds to two states of the entropy,  this  hints  a phase transition, while for positive  $c$ the entropy density takes the same value  for two different  $T$.

\section{Holographic model with two and three form fields in arbitrary dimension} \label{sec: section 3}
\subsection{The setup}

Now we turn to the second  model, which includes a scalar with its potential, one Maxwell field and one Kalb-Ramond field \cite{Kalb:1974yc}. The action in $D$ dimensions is given by:
\be
S = \int d^D x \, \sqrt{- g_D} \left(R- \cfrac{1}{2} \, \partial_{\mu} \phi \, \partial^{\,\mu} \phi - \cfrac{1}{4} \, f(\phi) F^2- \cfrac{1}{4} \, f_H(\phi)H^2 -V(\phi)\right),\label{action: two-mag-fields model}
\ee
where $F,\;H$ are  
two- and three-form fields, correspondingly, $g_D$ is the metric determinant, $\phi$ is a scalar field, $V(\phi)$ is its potential, $f(\phi),\; f_H(\phi)$ are the 
coupling functions associated with the corresponding 
fields.

The Einstein equations are defined by \eqref{EinEq} with the stress-energy tensor
\bea\label{SETm2}
T_{\mu\nu}&=&\frac{1}{2}\left(\partial_{\mu}\phi\partial_{\nu}\phi-\frac{1}{2}g_{\mu\nu}(\partial \phi)^2 -g_{\mu\nu} V(\phi)\right) + \frac{f(\phi)}{2}\left(F_{\mu\alpha}F_{\nu}^{\alpha}-\frac{1}{4}g_{\mu\nu}F^2\right)\\
&+&\frac{f_{H}(\phi)}{2}\left(\frac{3}{2}H_{\mu\alpha\beta}H^{\,\,\alpha\beta}_{\nu}-\frac{1}{4}g_{\mu\nu}H^2\right).\nonumber
\eea
The scalar field equation reads
\be\label{dilaton EOM: general2}
\Box \phi= \cfrac{1}{4}\cfrac{\partial f}{\partial\phi}F^2 + \cfrac{1}{4}\cfrac{\partial f_H}{\partial\phi}H^2 + \cfrac{\partial V}{\partial\phi}, \quad \Box=\frac{1}{\sqrt{|g|}}\partial_{\mu}\left(g^{\mu\nu}\sqrt{|g|}\partial_{\nu}\right).
\ee

The Bianchi identities for the form fields $B$ and $F$ can be represented as
\bea
\partial_\nu(\sqrt{-g}f(\phi)F^{\mu\nu})&\equiv&0\,,\\
\partial_{\nu}(\sqrt{-g}f_H(\phi)H^{\mu\nu\lambda})&\equiv& 0\,.
\eea
Note that as in the first model from Sect. \ref{sec: section 2}, the form fields $B$ and $F$ do not give any new independent equations to the system.

We consider the following ansatz for the $D$-dimensional metric:
\be
ds^2 = \cfrac{L^2b(z)}{z^2}\left(-g(z)dt^2 + \sum_{i=1}^d dx_i^2 + \sum_{j=1}^3 g_{y_jy_j}dy_j^2 + \cfrac{dz^2}{g(z)}\right),\label{D-metric: 2-mag-fields general}
\ee
where $d = D-5$, $b(z)$ is a warp factor, $g(z)$ is the blackening function and $L$ is the $AdS$ radius. The ``$y$''-coordinates correspond with the anisotropic part of the metric  \eqref{D-metric: 2-mag-fields general}.

We also suppose that the  scalar field depends only on the radial coordinate $z$
\be
\phi = \phi(z).\label{dilaton: 2-mag-fields model}
\ee
For both Maxwell and Kalb-Ramond fields we use a magnetic ansatz, i.e. the field strengths take the form
\bea
F &=& q\,dy_2\wedge dy_3,\label{F: 2-mag-fields model}\\
H &=& q_H\,dy_1\wedge dy_2\wedge dy_3\label{B: 2-mag-fields-model},
\eea
where $q$ and $q_H$ are some constant parameters. 

Note that in $D=5$, Hodge duality allows us to represent the 3-form in terms of 2-form $G$, such that $H=*G$. In other words, one can reformulate model \eqref{action: two-mag-fields model} in terms of two Maxwell fields and a dilaton with potential. In this dual picture, our model \eqref{action: two-mag-fields model} reduces to the one studied in the previous section.

\subsection{The equations of motion}

In this subsection we present the equations of motion to \eqref{action: two-mag-fields model} with a certain ansatz for the metric and fields.
We are interested in the following ansatz for the metric \eqref{D-metric: 2-mag-fields general} :
\be
ds^2 = \cfrac{L^2 \, b(z)}{z^2} \left(-g(z) dt^2 + \sum_{i=1}^d dx_i^2 + \cfrac{e^{c_Bz^2}dy_1^2 + dy_2^2 + dy_3^2 }{(z/L)^{\frac{2}{\nu} -2} } + \cfrac{dz^2}{g(z)} \right),\label{D-metric: 2-mag-fields}
\ee
where $d = D-5$  and $c_B$ is a parameter.
Taking into account \eqref{dilaton: 2-mag-fields model}, \eqref{F: 2-mag-fields model},\eqref{B: 2-mag-fields-model} and \eqref{D-metric: 2-mag-fields} and doing some algebra, the equations of motion \eqref{EinEq} with \eqref{SETm2}  are brought to the following form
\bea
& & 6c_B + 2c_B^2z^2  + \cfrac{6}{z^2\nu^2} - \cfrac{4 c_B}{\nu} - \cfrac{6}{z^2\nu} +\left(\cfrac{D}{2} - 1\right)\left(\cfrac{4b'}{zb}- \cfrac{3b'^2}{b^2}+ \cfrac{2b''}{b}\right) + \phi'(z)^2 = 0,\,\label{dilaton equation: 2-mag-fields model}\;\\
& &g'' + g'\left(\cfrac{b'}{b}\left(\cfrac{D}{2} - 1\right) + zc_B +\cfrac{1}{z}\left(5-D-\cfrac{3}{\nu}\right)\right) = 0\,,\label{g equation: 2-mag-fields model} \qquad\qquad \\
& & \cfrac{g'}{g} + \cfrac{b'}{b}\left(\cfrac{D}{2} - 1\right) + zc_B +\cfrac{1}{z}\left(6 - D -\cfrac{3}{\nu}\right) - \cfrac{q^2f(\phi)}{2zbgc_B}\left(\cfrac{z}{L}\right)^{\frac{4}{\nu}-2} = 0\,\label{f equation: 2-mag-fields model}, \\
& &\cfrac{g'}{g}\left(1 - \cfrac{1}{\nu} + z^2c_B\right) + \cfrac{b'}{b}\left(\cfrac{D}{2}-1\right)\left(z^2c_B- \cfrac{1}{\nu} + 1\right) + zc_B\left(7 - D + z^2c_B - \cfrac{4}{\nu}\right) + \nn \\
&&+\cfrac{1}{z}\left(4 - D + \cfrac{1}{\nu}\left(D-7\right) + \cfrac{3}{\nu^2}\right) + \cfrac{3ze^{-c_Bz^2}q_H^2f_H(\phi)}{2b^2g}\left(\cfrac{z}{L}\right)^{\frac{6}{\nu}-2} = 0\,\label{fB equation: 2-mag-fields model},\\
&&\cfrac{g''}{g} + \cfrac{g'}{g}\left(3zc_B +\frac{1}{z}\left(13 - 3D - \frac{7}{\nu}\right) + \cfrac{3b'}{b}\left(\frac{D}{2} - 1\right)\right) + \cfrac{b''}{b}(D-2) + \cfrac{b'^2}{b^2}\left(\frac{D^2}{2} \right. -  \nn \\
&&\left. -3D + 4\right) + \cfrac{b'}{b}\left(2zc_B(D - 2) -\cfrac{5}{z\nu}\left(D-2\right) + \cfrac{1}{z}\left(13D-2D^2 - 18\right)\right) + \cfrac{2L^2bV(\phi)}{z^2g} + \nn \\
&&+ 2c_B\left(10-2D - \cfrac{5}{\nu} + z^2c_B\right) +\cfrac{2}{z^2}\left(16-8D + D^2 + \cfrac{1}{\nu}\left(5D - 20\right) + \cfrac{6}{\nu^2}\right) = 0.\label{V equation: 2-mag-fields model}
\eea
The scalar  field equation according to \eqref{dilaton EOM: general2} with the metric \eqref{D-metric: 2-mag-fields} comes to:

\bea\label{scalfeqmod2}
\phi'' &+& \phi'\left(\cfrac{g'}{g} + \cfrac{b'}{b}\left(\cfrac{D}{2} - 1\right) + zc_B + \cfrac{1}{z}\left(5 - D - \cfrac{3}{\nu}\right)\right) - \cfrac{L^2b}{z^2g}\cfrac{\partial V}{\partial\phi} - \nn \\
&-&\cfrac{q^2}{2bg}\left(\cfrac{z}{L}\right)^{\frac{4}{\nu} - 2}\cfrac{\partial f}{\partial\phi} - \cfrac{3e^{-c_Bz^2}q^2_H}{2b^2g}\left(\cfrac{z}{L}\right)^{\frac{6}{\nu} - 2}\cfrac{\partial f_H}{\partial\phi}= 0.
\eea

\subsection{Black brane solutions with $b=e^{-cz^2}$}\label{subsec: section 3.4}

Now  we find  black brane solutions to \eqref{dilaton equation: 2-mag-fields model}-\eqref{scalfeqmod2} with the metric \eqref{D-metric: 2-mag-fields} in arbitrary $D$.
A solution for the blackening function $g(z)$ can be derived from \eqref{g equation: 2-mag-fields model} taking into account \eqref{boundary conditions}. Thus, we obtain:
\be
g(z) = 1 - \cfrac{\int\limits_{0}^z\cfrac{t^{ D+\frac{3}{\nu}-5 }e^{-\frac{t^2}{2}c_B  } }{b^{ \frac{D}{2}-1 } }dt}{\int\limits_{0}^{z_h}\cfrac{t^{ D+\frac{3}{\nu}-5 }e^{-\frac{t^2}{2}c_B  } }{b^{ \frac{D}{2}-1 } }dt}.
\ee
Motivated by the applications of the model \eqref{action: two-mag-fields model} to holographic studies of magnetic catalysis \cite{Arefeva:2023jjh}, we choose the warp factor to be $b(z) = e^{-cz^2}$, where $c$ is a constant parameter.   The solution for the blackening function $g$ with  \eqref{boundary conditions} takes the form
\be
g(z) = 1 - \cfrac{\int_0^z t^{D+\frac{3}{\nu} -5 } e^{-\frac{t^2}{2}\kappa}dt }{\int_{0}^{z_h} t^{D+\frac{3}{\nu} -5 }e^{-\frac{t^2}{2}\kappa } dt } = 1-\cfrac{\gamma(\alpha, z^2\kappa)}{\gamma(\alpha, z^2_h\kappa)} = \cfrac{\Gamma(\alpha, z^2_h\kappa) - \Gamma(\alpha, z^2\kappa) }{\Gamma(\alpha, z^2_h\kappa) - \Gamma(\alpha)},\label{blackening func solution b = e^c_Bz^2: 2-mag-fields model}
\ee
where we define
\bea\label{alphakappa}
\alpha = \cfrac{1}{2}\left(D+\cfrac{3}{\nu} - 4\right)\,,\quad \kappa = \cfrac{1}{2}\left(c_B - c(D-2)\right)
\eea
and $\gamma(x, y)$ is a lower incomplete gamma-function, $\Gamma(x, y)$ is an upper incomplete gamma-function.

Using \eqref{dilaton equation: 2-mag-fields model} the scalar field can be represented in the following form 
\be
\phi = \pm \int\sqrt{\cfrac{a}{z^2} -bz^2 + k }\;dz + \phi_0,
\ee
where
\be
a = \cfrac{6}{\nu}\left(1 - \cfrac{1}{\nu}\right),\quad
b = 2(c_B^2 - c^2(D-2)),\quad
k = \cfrac{4c_B}{\nu} - 6(c_B - c(D-2)).
\ee
Then the solution for $\phi$ is given by
\bea
\phi = \pm \cfrac{1}{2\sqrt{b}}\left[k\,\arctan\left({\cfrac{\sqrt{b}z^2}{\sqrt{a + kz^2 - bz^4} - \sqrt{a}}}\right) \right. + \sqrt{b}\left(\sqrt{a + kz^2 - bz^4} \right. \nn \\
\left.\left. - \sqrt{a}\log{z^2} + \sqrt{a}\log{\left(-2a - kz^2 + 2\sqrt{a}\sqrt{a + kz^2 - bz^4}\right)}  \right) \right] + \phi_0.
\eea
Note that near the boundary, as $z$ going to $0$, the scalar field is
\be
\phi \approx \pm \left(\cfrac{\sqrt{a}}{2} + \cfrac{k}{2\sqrt{b}}\arctan\left(\cfrac{2\sqrt{ab}}{k}\right) - \cfrac{\sqrt{a}}{2}\log\left(\cfrac{-4a}{(4ab + k^2)z^2}\right) \right) + \phi_0.
\ee


The coupling function $f$ can be found from eq.\eqref{f equation: 2-mag-fields model}
\be\label{ffuncmod2der}
f = \cfrac{2g(z)}{q^2}e^{-cz^2}\left(\cfrac{z}{L}\right)^{2-\frac{4}{\nu}}\left[c_Bz\cfrac{g'}{g} -z^2cc_B(D-2)- \cfrac{3c_B}{\nu} + c^2_Bz^2 - c_BD + 6c_B\right].
\ee
Plugging the blackening function $g$ \eqref{blackening func solution b = e^c_Bz^2: 2-mag-fields model} into \eqref{ffuncmod2der}, we get for $f$ 
\bea
f &=& \cfrac{2}{q^2}e^{-cz^2}\left(\cfrac{z}{L}\right)^{2-\frac{4}{\nu}}\cfrac{ \Gamma(\alpha, z^2_h\kappa) - \Gamma(\alpha, z^2\kappa) }{\Gamma(\alpha, z^2_h\kappa) - \Gamma(\alpha)}  \left[  \cfrac{  2c_B e^{-z^2\kappa  } \kappa^\alpha z^{2\alpha-1}} {\Gamma(\alpha, z^2_h\kappa) - \Gamma(\alpha, z^2\kappa)  } + 2z^2c_B\kappa - \cfrac{3c_B}{\nu} + \right.\nn \\
&& \left. + c^2_Bz^2 - c_BD + 6c_B\right]. 
\eea
Similarly, we find the second coupling function $f_H$ from \eqref{fB equation: 2-mag-fields model}
\bea
f_H &=& \cfrac{2g}{3q_H^2}e^{-z^2(2c - c_B)}\left(\cfrac{z}{L}\right)^{2-\frac{6}{\nu}}\left[\left(7-\cfrac{3}{\nu} -D\right)\left(\cfrac{1}{z^2\nu} - c_B\right) + \cfrac{g'}{zg}\left(\cfrac{1}{\nu} - {1} - c_Bz^2\right) \right. + \nn \\
&+& \left.\cfrac{c_B}{\nu} + \cfrac{D}{z^2} - \cfrac{4}{z^2} + c(D-2)\left(1-\cfrac{1}{\nu}\right)\right].
\eea
Utilizing the blackening function $g$~\eqref{blackening func solution b = e^c_Bz^2: 2-mag-fields model}, we obtain the following expression for $f_H$:
\bea
f_H&=& \cfrac{2}{3q^2_H}e^{-z^2(2c-c_B)}\left(\cfrac{z}{L}\right)^{2-\frac{6}{\nu}}\cfrac{ \Gamma(\alpha, z^2_h\kappa) - \Gamma(\alpha, z^2\kappa) }{\Gamma(\alpha, z^2_h\kappa) - \Gamma(\alpha)}\left[ \cfrac{c_B}{\nu} + \cfrac{D}{z^2} - \cfrac{4}{z^2} + c(D-2)\left(1-\cfrac{1}{\nu}\right) + \right.\nn \\
&+& \left. \cfrac{ 2e^{-z^2\kappa}\kappa^\alpha z^{2\alpha - 2} }{\Gamma(\alpha, z^2_h\kappa) - \Gamma(\alpha, z^2\kappa)}\left(\cfrac{1}{\nu} -1 - c_Bz^2\right) + \left(7 - \cfrac{3}{\nu} - D\right)\left(\cfrac{1}{z^2\nu} - c_B\right)\right].
\eea
Finally, the scalar potential
 is provided by the combination of equations \eqref{g equation: 2-mag-fields model} and \eqref{V equation: 2-mag-fields model}:
\bea
V &&= g\,e^{cz^2}\left(\cfrac{z}{L}\right)^2\left\{2(D-5)(c_B - c(D-2))-\left(\cfrac{D-4}{z}\right)^2 - z^2(c_B - c(D-2))^2 -\cfrac{6}{z^2\nu^2} - \right. \nn \\
&&- \cfrac{5}{z^2\nu}(D-4) + \cfrac{5}{\nu}(c_B - c(D-2))+ \left. \cfrac{g'}{g}\left[\cfrac{D-4+\frac{2}{\nu} }{z} - z(c_B - c(D-2))\right]  \right\} .
\eea
Substituting $g$ \eqref{blackening func solution b = e^c_Bz^2: 2-mag-fields model} into the latter equation, we find the expression for the potential 
\bea
V&& = \cfrac{ \Gamma(\alpha, z^2_h\kappa) - \Gamma(\alpha, z^2\kappa) }{ \Gamma(\alpha, z^2_h\kappa) - \Gamma(\alpha) }e^{cz^2}\left(\frac{z}{L}\right)^2\left\{ 2(D-5)(c_B - c(D-2))-\left(\frac{D-4}{z}\right)^2 - \cfrac{6}{z^2\nu^2}  - \right. \nn \\
&& \left. - z^2(c_B - c(D-2))^2 - \cfrac{5}{z^2\nu}(D-4) + \cfrac{5}{\nu}(c_B - c(D-2)) + \right. \nn \\
&&+\left. \cfrac{ 2\kappa^\alpha z^{2\alpha-1}e^{-z^2\kappa} }{\Gamma(\alpha, z^2_h\kappa) - \Gamma(\alpha, z^2\kappa)}\left[\cfrac{D-4+\frac{2}{\nu} }{z} - z(c_B - c(D-2))\right]\right\}.
\eea

The Hawking temperature to the black brane solution with \eqref{blackening func solution b = e^c_Bz^2: 2-mag-fields model} can be found as for the previous model \eqref{hawking temperature formula} :
\be\label{hawktempmod2wf}
T(z_h) = \cfrac{1}{2\pi}\left|\cfrac{e^{z^2_h\kappa}z^{2\alpha - 1}_h \kappa^\alpha}{\Gamma(\alpha, z^2_h\kappa) - \Gamma(\alpha)}\right|.
\ee

The entropy density can be calculated  using  \eqref{entropy formula} with our ansatz \eqref{D-metric: 2-mag-fields} and $b=e^{-cz^{2}}$, thus we have
\be\label{entwfmod2}
s(z_h) =
\cfrac{1}{4}\left(\cfrac{z_h}{L}\right)^{5-D-\frac{3}{\nu} }e^{\frac{z^2_h}{2}(c_B - c(D-2)) }
\ee
or in terms of $\alpha$ and $\kappa$ \eqref{alphakappa}
\be\label{entalphakappamod2}
s(z_h)= \cfrac{1}{4}\left(\cfrac{L}{z_h}\right)^{2\alpha - 1} e^{\kappa z_h^2}.
\ee

\begin{figure}[H]
\centering
\includegraphics[scale = 0.86]{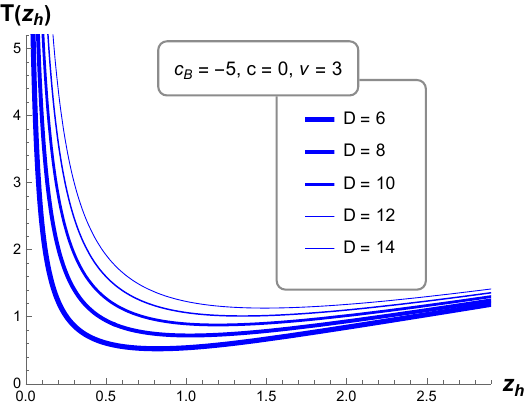}
\includegraphics[scale = 0.86]{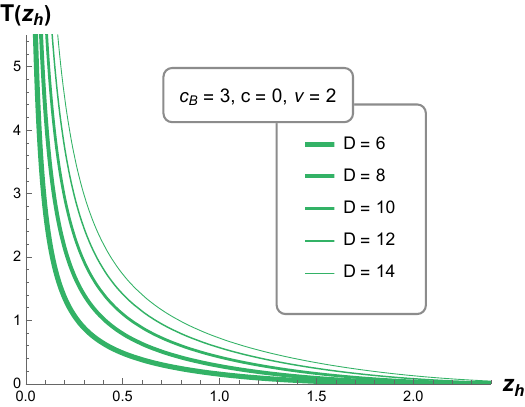}\\
(A)\hspace{200pt} (B)
\includegraphics[scale = 0.86]{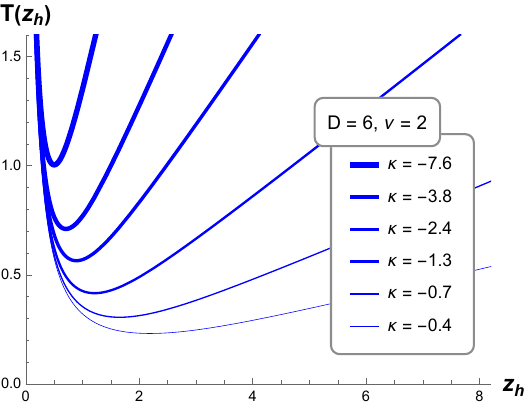}
\includegraphics[scale = 0.86]{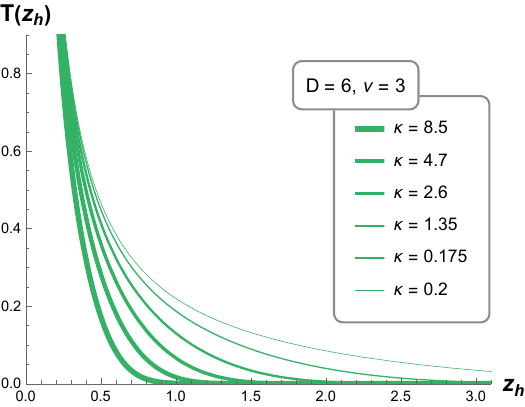}\\
(C) \hspace{200pt} (D)
\caption{Hawking temperature $T$ on $z_h$ for the black brane solutions with $b=e^{-cz^2}$ for various $D$, $c_B$, $c$, $\nu$. In Figs. (A) and (B) we fix $c_B$, $c$ and $\nu$ and vary $D$. In Figs. (C) and (D) we fix $D$ and $\nu$ and vary $\kappa$ defined in \eqref{alphakappa}.
}\label{fig:Tzhmod2}
\end{figure}

\begin{figure}[H]
    \centering
    \includegraphics[scale = 0.86]{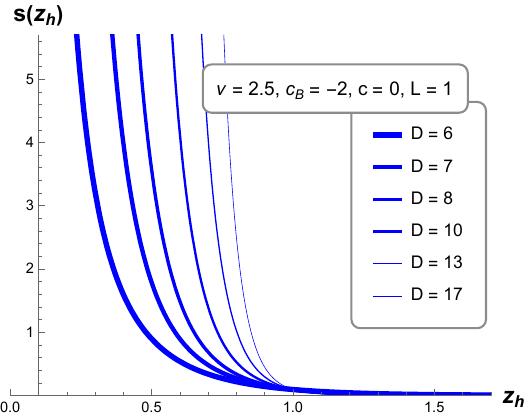}
    \includegraphics[scale = 0.86]{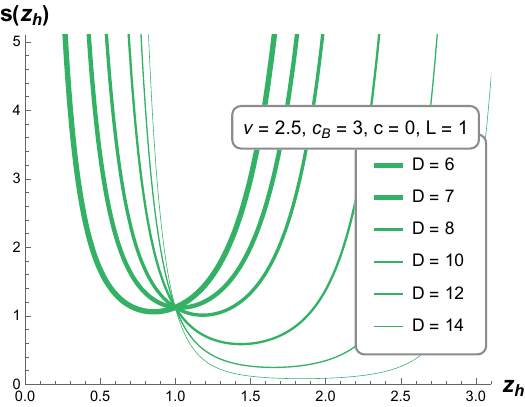}
    \includegraphics[scale = 0.86]{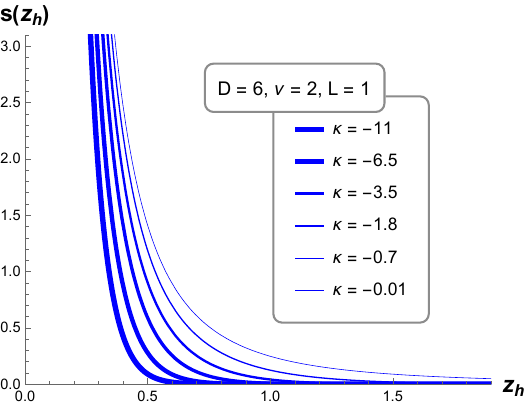}
    \includegraphics[scale = 0.86]{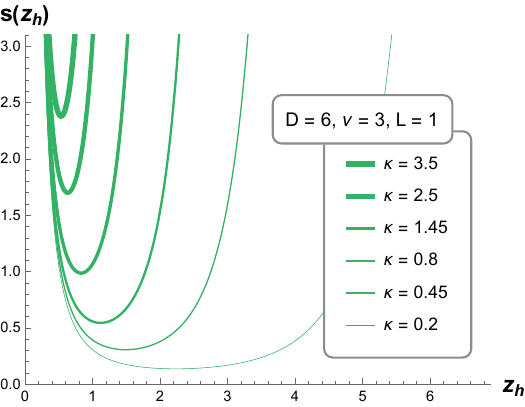}
    \caption{Dependence of the entropy density on $z_h$ for the black brane solutions with $b=e^{-cz^2}$ for different sets of parameters. In Figs. (A) and (B)  we fix $\nu=2.5$, $c_B$ and $c=0$ and vary $D$, while for (C) and (D) we keep fixed $D=6$, $\nu$ and vary $\kappa$. 
    For all figures we put $L=1$.}
    \label{fig:szhmod2}
\end{figure}

 Equations \eqref{hawktempmod2wf} and \eqref{entalphakappamod2} demonstrate that the Hawking temperature and the entropy density depend on the parameters $\alpha$ and $\kappa$, rather than on $\nu$, $c$, and $c_B$ individually. Hereafter, and without loss of generality, we set $c = 0$ and treat $\kappa$ as a free parameter in our plots. In Figs.~\ref{fig:Tzhmod2} and ~\ref{fig:szhmod2} we show  the temperature and entropy density as functions of $z_h$ for different  $D$ and $c_B$. In Figs.~\ref{fig:Tzhmod2} (A) and (B) we vary $D$ with fixed $c_B$ and $\nu$. With $c = 0$, the parameter $\kappa$ remains fixed; consequently, any variation in $D$ affects only $\alpha$. In Figs. (C) and (D) we fix the parameter $\alpha$ \eqref{alphakappa} for certain $D$ and $\nu$ and vary $\kappa$. In Figs. (A) and (C) the blue curves correspond to the case $\kappa < 0$, for which  we see that the Hawking temperature is non-monotonic and has a minimum. In Figs. (B) and (D) we show  the dependence of the Hawking temperature on $z_h$ for $\kappa>0$ by green. In this case $T(z_h)$ is a monotonically decreasing function.

In Fig.~\ref{fig:szhmod2} we depict the entropy density on $z_h$ for different sets of parameters. In Figs. (A) and (B) we vary $D$ and keep $\kappa$ fixed by setting $c = 0$ and fixing $c_B$ and $\nu$. For Figs. (C) and (D) we vary $\kappa$ with fixed $D$ and $\nu$.
In (A) and (C) the entropy density on $z_h$ decreases monotonically, while for (B) and (D) the function $s(z_h)$ is non-monotonic.

\begin{figure}[H]
    \centering
    \includegraphics[scale = 0.86]{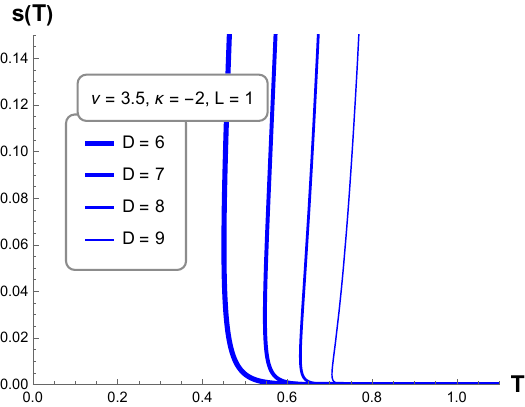}
    \includegraphics[scale = 0.86]{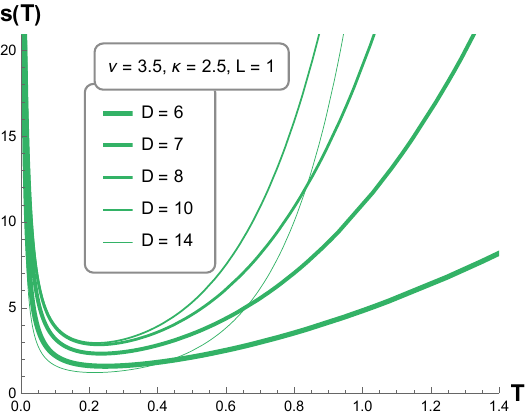}\\
    \includegraphics[scale = 0.96]{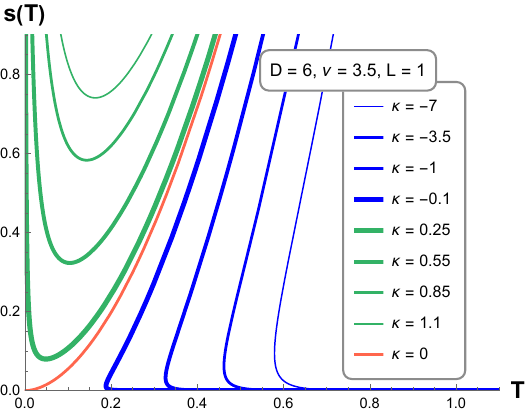}
    \caption{Dependence of the entropy density on the Hawking temperature for black brane solutions with $b=e^{-cz^2}$ for different sets of parameters. For Figs. (A) and (B) we  fix $\nu$, $\kappa$ and vary $D$. The case $\kappa < 0$ is shown in Fig. (A), while Fig. (B) corresponds to $\kappa > 0$. In Fig. (C) we fix $D$, $\nu$ and vary the parameter $\kappa$.  For all figures we set $c = 0$, $L=1$.}
    \label{fig: s(T)}
\end{figure}

In Fig.~\ref{fig: s(T)} we depict the behaviour of the entropy density as a function of $T$ for black brane solutions with $b=e^{cz^2}$. In Figs. ~\ref{fig: s(T)} (A) and (B) we vary $D$ fixing $\nu$ and $\kappa$. From Fig.~\ref{fig: s(T)} we see that for any dimension for both negative and positive $\kappa$    the entropy density  has a non-monotonic dependence on the temperature $T$, $s(T)$ is multivalued in Fig. (A) and has a minimum in (B) for $\kappa < 0$ ($c_B < 0$), for this case two different values of the temperature are related to a certain value of $s$. This can be also observed from Fig. (C), for which  we fix $D$, $\nu$ and vary $\kappa$. The only case for which the dependence $s$ on $T$ is monotonic is $\kappa=0$, for which we see that  $s\to  0$ as $T\to 0$. Note that  all figures 
are presented for $c = 0$ and different $c_B$, which in turn corresponds to different $\kappa$.

\subsubsection{Special case of the black branes with $c= c_B/(D-2)$}\label{subsec: section 3.5}

Here we consider a special case of the solutions with the Gaussian warped factor   $b=e^{-c z^2}$ with a constraint between parameters $c$ and $c_B$, namely $c = \cfrac{c_B}{D-2}$, which is equivalent to $\kappa = 0$ due to \eqref{alphakappa}. Under this constraint, the equations of motion can be simplified to
\bea\label{gconstrmod2}
&&g'' - \cfrac{g'}{z}\left(D-5+\cfrac{3}{\nu}\right) = 0,\\ \label{phiconstrmod2}
&&\cfrac{2c_B^2z^2}{D-2}(D-3) -\cfrac{4c_B}{\nu} + \cfrac{6}{z^2\nu^2} - \cfrac{6}{z^2\nu} + \phi'^2 = 0.
\eea
From \eqref{gconstrmod2} the solution for the blackening  function reads
\bea\label{blackfuncmod2constr}
&&g(z) = 1 - \left(\cfrac{z}{z_h}\right)^{D-4+\frac{3}{\nu} }.
\eea
Despite the background has a non-trivial warped factor $b$, the scalar field $\phi$ can be easily found from \eqref{phiconstrmod2}
\be
\phi = \pm \int\sqrt{\cfrac{6}{z^2\nu}\left(1 - \cfrac{1}{\nu}\right) + \cfrac{4c_B}{\nu} - \cfrac{2c_B^2(D-3)z^2}{D-2} }\, dz + \phi_0.
\ee

The coupling functions $f(z)$ and $f_H(z)$ can be derived from  eqs.\eqref{f equation: 2-mag-fields model}-\eqref{fB equation: 2-mag-fields model},  correspondingly,
\bea
&&f = \cfrac{2c_B}{q^2}e^{-\frac{c_Bz^2}{D-2}  }\left(\cfrac{z}{L}\right)^{2-\frac{4}{\nu}}\left[6-D-\cfrac{3}{\nu}-2\left(\cfrac{z}{z_h}\right)^{D-4+\frac{3}{\nu}}\right],\\
&&f_H = -\cfrac{2e^{\frac{c_Bz^2}{D-2} (D-4)}}{3L^2q^2_H\nu}\left(\cfrac{z}{L}\right)^{-\frac{6}{\nu}}\left[(1-\nu-\nu c_Bz^2)\left(D-4+\frac{3}{\nu}\right)\left(\cfrac{z}{z_h}\right)^{D-4+\frac{3}{\nu} } \right. +\nn \\ 
&& +\left. \left(1 - \left(\cfrac{z}{z_h}\right)^{D-4+\frac{3}{\nu}}\right)\left(\cfrac{6}{\nu}-14+2D - 6c_Bz^2 - 2\nu(D-4 + (D-6)c_Bz^2) \right)\right].\qquad\qquad
\eea
Finally, the potential is defined through \eqref{V equation: 2-mag-fields model} and has a quite simple form
\be
V= -e^{-\frac{c_Bz^2}{D-2}}\cfrac{(3+(D-4)\nu)(2+(D-4)\nu)}{L^2\nu^2}.
\ee

The Hawking temperature  \eqref{hawktempmod2wf} and the entropy density \eqref{entwfmod2} of the black brane solutions with \eqref{blackfuncmod2constr} take the form 
\be
T(z_h) = \left|\cfrac{D-4+\frac{3}{\nu}}{4\pi z_h}\right|,\quad s(z_h) = \cfrac{1}{4}\left(\cfrac{z_h}{L}\right)^{5-D-\frac{3}{\nu}}.
\ee
Thus, the black brane solutions with the constraint $c=\frac{c_B}{D-2}$ the dependence of the entropy density of the temperature has a power law
\be\label{sTmod2const}
s(T) = \cfrac{1}{4}\left(\cfrac{D-4+\frac{3}{\nu}}{4\pi L T}\right)^{5-D-\frac{3}{\nu}}.
\ee
One can see from \eqref{sTmod2const} that the third law of thermodynamics is also satisfied for $\nu \geqslant 1$ and $D\geqslant 5$.

\subsection{$5$-dimensional black brane solutions with  $b=1$}\label{subsec: section 3.3}

In this subsection, we focus on 5-dimensional  black brane solutions with an ansatz for the metric such that $b=1$ and $c_B = 0$, i.e.:
\be\label{metricansatz5d}
ds^2 = \cfrac{L^2}{z^2}\left(-g(z)dt^2 + \left(\cfrac{z}{L}\right)^{2-\frac{2}{\nu}}(dy_1^2 + dy_2^2 + dy_3^2) + \cfrac{dz^2}{g(z)}\right).
\ee
In this case the equations of motion take the form
\bea
&&\phi'^2 = \cfrac{6}{z^2\nu}\left(1 - \cfrac{1}{\nu}\right),\label{phi equation two-mag-fields model special case}\\
&&f(\phi) \equiv 0,\\
&&\cfrac{g''}{g} -\cfrac{g'}{g}\left(\cfrac{7}{z\nu} + \cfrac{2}{z}\right) + \cfrac{2}{z^2}\left(1 + \cfrac{5}{\nu} + \cfrac{6}{\nu^2}\right) + \cfrac{2L^2}{z^2g}V = 0\,,\label{g equation two-mag-fields model special case}\\
&&\cfrac{g''}{g} - \cfrac{g'}{g}\left(\cfrac{2}{z} + \cfrac{1}{z\nu}\right) - 3L^2\left(\cfrac{z}{L}\right)^{\frac{6}{\nu} - 2}\cfrac{q_H^2}{g}f_H + \cfrac{2}{z^2}\left(1 + \cfrac{2}{\nu} - \cfrac{3}{\nu^2}\right) = 0.\,\label{fb equation two-mag-fields model special case}
\eea
A linear combination of the Einstein equations leads to the vanishing of the Maxwell field because of the field configuration.

We are interested in two particular cases: {\bf 1)} $V=\Lambda$ and {\bf 2)} $V=0$. For the first case the  blackening function $g(z)$ with  \eqref{boundary conditions} is given by :
\bea\label{blackfunc5d2mod}
g(z) = 1 + C\,z^{1+\frac{3}{\nu}} - \left(\cfrac{z}{z_h}\right)^{2+\frac{4}{\nu}}\left(1 + C\,z_h^{1+\frac{3}{\nu}}\right),
\eea
where $C$ is an arbitrary constant.

The  parameters are related by the following constraint 
\be\label{constraints5dmod}
L^2\Lambda = -\frac{{(\nu+2)(\nu+3)}}{\nu^2}.
\ee

The scalar field can found from eq.~\eqref{phi equation two-mag-fields model special case}, so we obtain 
\be
\phi = \phi_0 + \lambda_5\log z^{-\frac{6}{\nu}}, \quad\lambda_5 = \pm\sqrt{\cfrac{\nu - 1}{6}}\label{phi solution el-mag model special case}.
\ee
Note that for $\nu =1$ the scalar field $\phi$ is constant.
The coupling  function $f_H$ can be represented in the following form
\be
f_H = \cfrac{ 2L^{\frac{6}{\nu} - 4} }{q^2_H\nu^2}z^{-\frac{6}{\nu} } \left[\cfrac{1}{3}(\nu-1)(\nu+3) - (\nu+1)\left(\cfrac{z}{z_h}\right)^{2+\frac{4}{\nu}}\left(1 + {C}\,z_h^{1+\frac{3}{\nu}}\right) \right].\label{fHsol5Dmod}
\ee

The Hawking temperature of the solution with the metric \eqref{metricansatz5d} and \eqref{blackfunc5d2mod} is given by
\be\label{Tzh5dmod2}
T(z_h) = \cfrac{\left|C\,\left(1+\frac{1}{\nu}\right)z_h^{1+\frac{3}{\nu}} + 2+\frac{4}{\nu}\right|}{4\pi z_h}.
\ee

The entropy density of the black brane solution with $b=1$ reads
\be\label{szh5dmod2}
s(z_h)=\cfrac{1}{4}\left(\cfrac{z_h}{L}\right)^{-\frac{3}{\nu}}.
\ee

\begin{figure}[H]
\centering
\includegraphics[scale = 1.]{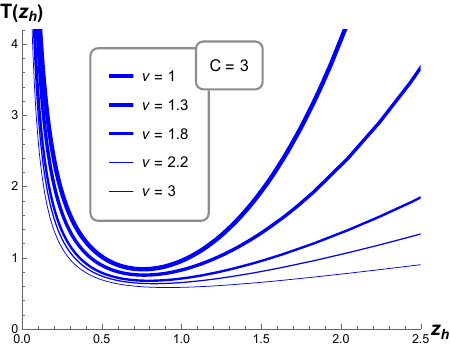}
\includegraphics[scale = 1.]{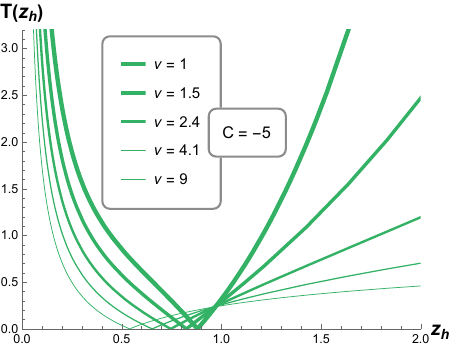}
(A) \hspace{200pt} (B) \\
\includegraphics[scale = 1.]{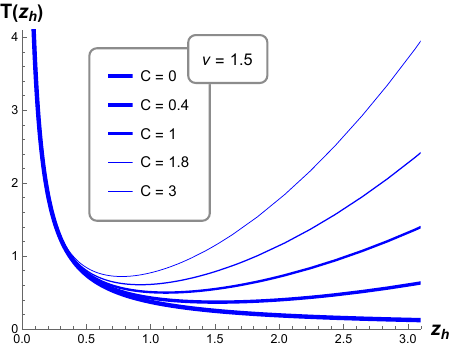}
\includegraphics[scale = 1.]{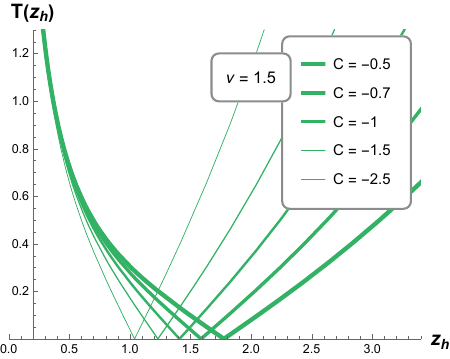}\\ 
(C) \hspace{160pt} (D) \\
\caption{Hawking temperature as a function $z_h$ for the 5-dimensional black brane with the warp factor $b = 1$ and $V = \Lambda$. In Figs. (A) and (B) we fix $C$  and vary $\nu$, for (C) and (D) we fix $\nu$ and vary $C$,   defined  in \eqref{blackfunc5d2mod}. In  Figs. (A) and (C) we depict the behavior of the temperature for $C\geqslant 0$, and figures (B) and (D) - for $C \leqslant 0$. 
}
\label{fig:Tzh350}
\end{figure}

In Fig.~\ref{fig:Tzh350} we depict the dependence of the Hawking temperature on $z_h$ for different sets of parameters. In Figs.~(A) and (B) we plot  $T$ on $z_h$ for different $\nu$ with fixed $C$ (positive and negative, correspondingly).  In Figs.~(C) and (D)  we plot $T(z_h)$ for fixed $\nu$ varying $C$. We observe that a certain value of the black brane temperature corresponds to two different values of $z_h$, excepting the case $C=0$. Moreover, in the case of negative $C$, the temperature of the black brane \eqref{Tzh5dmod2} reaches $0$ for certain non-zero $z_{h,0}$, i.e. the  horizon can be degenerate for this solution.

\begin{figure}[H]
\centering
\includegraphics[scale = 1.]{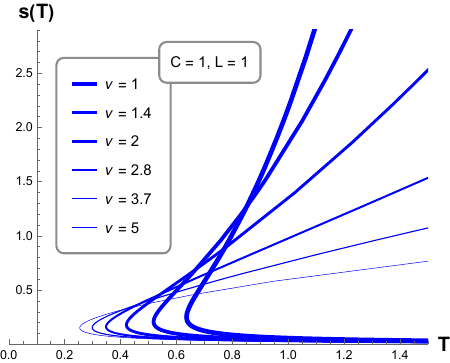}
\includegraphics[scale = 1.]{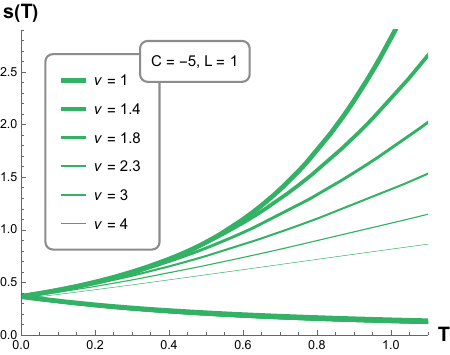}\\
(A)\hspace{200pt} (B)\\
\includegraphics[scale = 1.]{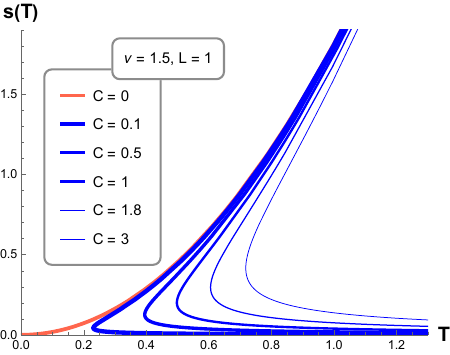}
\includegraphics[scale = 1.]{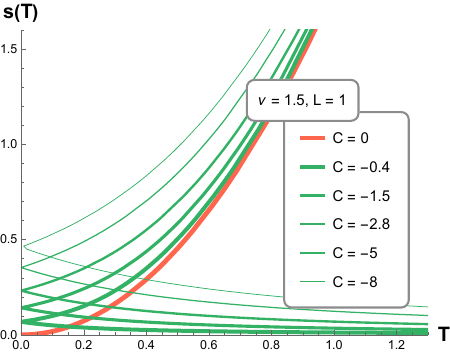}\\
(C)\hspace{200pt} (D)
\caption{Entropy density as a function of $T$ for the black brane solutions with \eqref{blackfunc5d2mod}. In Figs. (A) and (B) $C$ is fixed while $\nu$ varies. In Figs. (C) and (D) we vary $c$ and fix $\nu$. For all we set $L=1$. }
\label{fig:placeholder}
\end{figure}

In Fig.~\ref{fig:placeholder} we show  the dependence of the entropy density on the temperature.  We fix $C$ and vary $\nu$ in (A) and (B) and, in opposite, for (C) and (D) we fix $\nu$ and vary $C$. We see that the entropy density has a minimal value, above which for any $C$ and $\nu$ we have two values of the entropy density for a certain value of the temperature. The only case when the entropy density monotonically increases with $T$ is $C=0$.   In Figs. 7 (B) and (D) the lower branches of the green curves correspond to values $z_h < z_{h,0}$, the upper branches of the curves correspond to $z_h >z_{h,0}$.

Alternatively, we can fix the coupling function $f_H$ in the  following  form
\be
 f_H = e^{\mathrm{k}\phi}.
\ee 
Then from the  equations of motion \eqref{phi equation two-mag-fields model special case}-\eqref{fb equation two-mag-fields model special case} we find the constraints for the parameters 
\be
\mathrm{k}\lambda_{5}=1,\quad
 3L^{2-\frac{6}{\nu}}q_H^2= 2\left(1 + \cfrac{2}{\nu} - \cfrac{3}{\nu^2}\right),\,
\ee
with \eqref{constraints5dmod} and the constant $C$ is fixed as
\be
C = -\cfrac{1}{z_h^{1+\frac{3}{\nu}} }.
\ee


Now we turn to the second case, i.e. the vanishing potential $V = 0$. From the combination of equations \eqref{g equation two-mag-fields model special case} and \eqref{fb equation two-mag-fields model special case} we  get a solution for the blackening function
\be\label{blfuncmod2v0}
g =  1-\left(\cfrac{z}{z_h}\right)^{1+\frac{3}{\nu}}.
\ee



\begin{figure}[h!]
    \centering
    \includegraphics[scale = 1.]{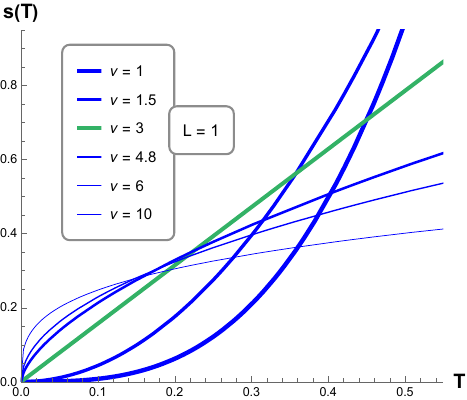}
    \caption{The entropy density of the 5-dimensional magnetic black brane \eqref{blfuncmod2v0} with  $V = 0$ for different $\nu$.} 
    \label{fig: s(T) two-mag-fields model V = 0 D = 5}
\end{figure}

It is interesting to note  that in this case the thermodynamics of the black brane solution matches with the thermodynamics of the 5-dimensional magnetic black brane \eqref{magneticb1}-\eqref{phisolmagbr5} from  Section \ref{sec: section 2}. The Hawking temperature and the dependence of the entropy density on the temperature are given, correspondingly,
\bea
T(z_h) = \cfrac{1+\frac{3}{\nu}}{4\pi z_h}, \qquad s(T) = \cfrac{1}{4}\left(\cfrac{1+\frac{3}{\nu}}{4\pi T}\right)^{-\frac{3}{\nu}}.
\eea
Comparing the latter  with \eqref{entropy on T: el-mag-fields model} we find that it coincides for $D=5$. We show the dependence of the entropy density as a function of $T$ in Fig.~\ref{fig: s(T) two-mag-fields model V = 0 D = 5}. We see that the behaviour of $s$ on $T$ differs for various $\nu$.  Particularly, for $\nu=3$ the entropy density depends linearly on $T$.

\section{Discussion}

In this work, we have constructed  families of black brane solutions with Lifshitz‑like asymptotics for two distinct holographic models in arbitrary spacetime dimensions $D$. The first model consists of a scalar field with a potential coupled to two Maxwell fields, admitting both electric and magnetic charges. The second model comprises a scalar field with a potential, a Maxwell field, and  a three-form field strength of a Kalb‑Ramond field. For each model, we derived exact solutions for the metric, scalar field, gauge fields, and coupling functions, incorporating general warp factors $b(z)$ (including Gaussian profiles) and anisotropic scaling characterized by the exponent $\nu$. Our results provide a natural generalization of previously known five‑dimensional anisotropic black brane solutions to arbitrary $D$.

A key difference between the two models lies in the degree of freedom count relative to the equations of motion. In the first model, particularly when both electric and magnetic fields are present, the system contains more unknown functions than independent equations. Consequently, the coupling functions $f_1$ and $f_2$ are not uniquely determined by the equations but are related to each other through the solution; additional input (such as a specific form for $f_1$) is required to fully specify the system. By contrast, the second model is fully determined: given the metric ansatz and the choice of warp factor $b(z)=e^{-cz^2}$, the equations of motion yield explicit closed‑form expressions for both coupling functions $f(\phi)$ and $f_H(\phi)$ without further assumptions. This makes the second model particularly tractable for holographic applications where precise knowledge of the matter couplings is important.

The thermodynamic analysis reveals a rich and parameter‑dependent behavior. For the simplest isotropic case ($b(z)=1$ and $\nu=1$), we recover standard AdS black branes. For anisotropic backgrounds with $b(z)=1$ and $\nu>1$, the entropy density obeys a power‑law relation $s(T)\propto T^{\alpha}$ with $\alpha>0$, so that $s\to 0$ as $T\to 0$. This is consistent with the third law of black hole thermodynamics. Including a non‑zero electric field modifies the temperature–entropy relation, but the third law remains satisfied provided the coupling function $f_1$ satisfies the condition $\kappa<-1$ (i.e., $2k\lambda>1$).

A more complex picture emerges when Gaussian warp factors are introduced. In the first model ($D=5$, $b=e^{cz^2}$), the entropy density as a function of temperature is monotonic and respects the third law only for $c=0$. For $c\neq0$, the $s(T)$ curves become non‑monotonic and may exhibit multivalued behavior, indicating the possibility of phase transitions and a violation of the third law. Remarkably, a similar pattern appears in the second model for $D>5$ with the warp factor $b=e^{-cz^2}$ and a non‑trivial parameter $\kappa$. Here, the entropy–temperature relation is non‑monotonic for both negative and positive $\kappa$, and only the special case $\kappa=0$ (which reduces to a power‑law form) yields $s\to0$ as $T\to0$. Moreover, the thermodynamic behavior of the $D$-dimensional second model for $\kappa\neq0$ mirrors that of the five‑dimensional first model with $c\neq0$, suggesting a certain universality in the way Gaussian deformations affect the third law.

 Within the holographic framework, these anisotropic backgrounds with nontrivial warp factors are relevant for describing strongly coupled systems with spatial anisotropy, such as those arising in heavy‑ion collisions (e.g., magnetic catalysis). The observed non‑monotonic entropy–temperature relations may signal phase transitions between different black brane branches, analogous to small/large black hole phase transitions in extended thermodynamics. The violation of the third law for certain parameter ranges indicates that such backgrounds cannot be obtained from a non‑extremal configuration by a finite physical process.

Several  directions for further research naturally follow from this work. First, it would be interesting to explore the stability of the constructed solutions under perturbations, particularly in the regimes where $s(T)$ is multivalued. Second, the holographic dual interpretation of the non‑monotonic thermodynamics deserves further investigation. Finally, the connection between the observed violations of the third law and the negative‑dimension Bose gas models mentioned in the Introduction suggests a deeper link between black hole thermodynamics and statistical mechanics that warrants further exploration.

In \cite{ARZ}, black brane solutions to the Einstein-dilaton-Maxwell models in $D=5$ and $D=6$ were constructed, and the null energy condition was explored. It would also be valuable to analyze the NEC for our solutions along the lines of \cite{ARZ}.

\section*{Acknowledgment}
 We are grateful to  Kristina Rannu,  Viktor Zlobin, Pavel Slepov and Igor Volovich for useful discussions.
The work of I.A. was performed at the Steklov  Mathematical Institute
 and supported by the Russian Science Foundation grand 24-11-00039.
\newpage
\appendix
\section{D-dimensional Einstein tensor }\label{sec: appendix A}
Let us derive explicit formulas for Einstein tensor in arbitrary dimensions using a particular ansatz.

We will consider diagonal metric depending only on holographic coordinate $z$:
\be
g_{\mu\nu} = diag\left(g_{00}(z),\ldots, g_{ii}(z),\ldots,g_{D-1D-1}(z)\right),\label{D-metric: general}
\ee
where $g_{00}\equiv g_{tt}$ and $g_{D-1D-1}\equiv g_{zz}$.

We now present expressions for the components of the Ricci tensor in terms of the components of the metric tensor \eqref{D-metric: general}
\bea
R_{zz} &=& {\cfrac{1}{4}\left(\cfrac{g'_{zz}}{g_{zz}}\right)\sum_{\gamma\neq z}\left(\cfrac{g'_{\gamma\gamma}}{g_{\gamma\gamma}}\right) + \cfrac{1}{4}\sum_{\gamma\neq z}\left(\cfrac{g'_{\gamma\gamma}}{g_{\gamma\gamma}}\right)^2 - \cfrac{1}{2}\sum_{\gamma\neq z}\left(\cfrac{g''_{\gamma\gamma}}{g_{\gamma\gamma}}\right)   },\label{D-dimensional ricci zz}\\
R_{\alpha\alpha} &=& { -\cfrac{1}{4}\left(\cfrac{g'_{\alpha\alpha}}{g_{zz}}\right)\sum_{\gamma\neq\alpha, z}\cfrac{g'_{\gamma\gamma}}{g_{\gamma\gamma}} + \cfrac{1}{4}\cfrac{g'_{\alpha\alpha}}{g_{zz}}\left(\cfrac{g'_{\alpha\alpha}}{g_{\alpha\alpha}} + \cfrac{g'_{zz}}{g_{zz}}\right) - \cfrac{1}{2}\cfrac{g''_{\alpha\alpha}}{g_{zz}},\;\;\alpha\neq z   }\label{D-dimensional ricci alpha alpha}
\eea
where we used $g^{\alpha\alpha} = \cfrac{1}{g_{\alpha\alpha}}$ since the metric tensor is diagonal. Therefore, using standard definition of Einstein tensor,
we obtain explicit formula for components of Einstein tensor in terms of metric tensor components:
\bea
G_{zz} &=& \cfrac{1}{8}\sum_{\alpha,\beta\neq z}\cfrac{g'_{\alpha\alpha}g'_{\beta\beta}}{g_{\alpha\alpha}g_{\beta\beta}} - \cfrac{1}{8}\sum_{\alpha\neq z}\left(\cfrac{g'_{\alpha\alpha}}{g_{\alpha\alpha}}\right)^2,\label{D-dimensional einstein tensor zz}\\
G_{\alpha\alpha} &=&  - \cfrac{1}{4}\cfrac{g_{\alpha\alpha}g'_{zz}}{g^2_{zz}}\sum_{\gamma\neq\alpha,z}\cfrac{g'_{\gamma\gamma}}{g_{\gamma\gamma}}- \cfrac{1}{4}\cfrac{g_{\alpha\alpha}}{g_{zz}}\sum_{\gamma\neq\alpha,z}\left(\cfrac{g'_{\gamma\gamma}}{g_{\gamma\gamma}}\right)^2+ \cfrac{1}{2}\cfrac{g_{\alpha\alpha}}{g_{zz}}\sum_{\gamma\neq\alpha,z}\cfrac{g''_{\gamma\gamma}}{g_{\gamma\gamma}}+\nn \\
&+& \cfrac{1}{4}\;\;
\cfrac{g_{\alpha\alpha}}{g_{zz}}\sum_{\substack{\lambda,\gamma\neq\alpha,z,\\ \lambda<\gamma}  }\cfrac{g'_{\lambda\lambda}g'_{\gamma\gamma}}{g_{\lambda\lambda}g_{\gamma\gamma}}.\label{D-dimensional einstein tensor alpah alpha}
\eea

\section{Stress-energy tensor structure}\label{sec: appendix B}
\subsection{Energy momentum tensor for the first ansatz}


The non-zero components of energy momentum tensor the for the first model with metric \eqref{D-metric: el-mag fields} are given by expressions:

\bea
\cfrac{T_{tt}}{g_{tt}} &=& -\cfrac{\phi'^2}{4b}z^2g - \cfrac{V}{2} -\cfrac{f_1}{4}A_t'^2\cfrac{z^4}{b^2} - \cfrac{1}{4}f_2\cfrac{q^2z^{\frac{4}{\nu}}}{b^2}\\
\cfrac{T_{zz}}{g_{zz}} &=& \cfrac{\phi'^2}{4b}z^2g-\cfrac{V}{2} - \cfrac{f_1}{4}A_t'^2\cfrac{z^4}{b^2} - \cfrac{1}{4}f_2\cfrac{q^2z^{\frac{4}{\nu}} }{b^2}\\
\cfrac{T_{x_ix_i}}{g_{x_ix_i}} &=& -\cfrac{\phi'^2}{4b}z^2g - \cfrac{V}{2} + \cfrac{f_1}{4}A_t'^2\cfrac{z^4}{b^2} - \cfrac{1}{4}f_2\cfrac{q^2z^{\frac{z}{\nu}} }{b^2},\quad i = \overline{1,d}\\
\cfrac{T_{y_jy_j}}{g_{y_jy_j}} &=& -\cfrac{\phi'^2}{4b}z^2g - \cfrac{V}{2} + \cfrac{f_1}{4}A_t'^2\cfrac{z^4}{b^2} + \cfrac{1}{4}f_2\cfrac{q^2z^{\frac{z}{\nu}} }{b^2},\quad j=\overline{1,2}
\eea

\subsection{Energy momentum tensor for the second ansatz}
Non-zero components of the second energy momentum tensor that is given by scalar field and two Maxwell fields - $F$ and $H$:
\bea
\cfrac{T_{tt}}{g_{tt}} &=& -\cfrac{\phi'^2}{4L^2b}z^2g - \cfrac{V}{2} - \cfrac{f}{4}\left(\cfrac{z}{L}\right)^\frac{4}{\nu}\cfrac{q^2}{b^2} - \cfrac{3}{4}f_H\left(\cfrac{z}{L}\right)^\frac{6}{\nu}e^{-c_Bz^2}\cfrac{q^2_H}{b^3}\\
\cfrac{T_{zz}}{g_{zz}} &=& \cfrac{\phi'^2}{4L^2b}z^2g - \cfrac{V}{2} - \cfrac{f}{4}\left(\cfrac{z}{L}\right)^\frac{4}{\nu}\cfrac{q^2}{b^2} - \cfrac{3}{4}f_H\left(\cfrac{z}{L}\right)^\frac{6}{\nu}e^{-c_Bz^2}\cfrac{q^2_H}{b^3}\\
\cfrac{T_{x_ix_i}}{g_{x_ix_i}} &=& -\cfrac{\phi'^2}{4L^2b}z^2g - \cfrac{V}{2}- \cfrac{f}{4}\left(\cfrac{z}{L}\right)^\frac{4}{\nu}\cfrac{q^2}{b^2} - \cfrac{3}{4}f_H\left(\cfrac{z}{L}\right)^\frac{6}{\nu}e^{-c_Bz^2}\cfrac{q^2_H}{b^3},\quad i = \overline{1, d}\\
\cfrac{T_{y_1y_1}}{g_{y_1y_1}} &=&  -\cfrac{\phi'^2}{4L^2b}z^2g - \cfrac{V}{2}- \cfrac{f}{4}\left(\cfrac{z}{L}\right)^\frac{4}{\nu}\cfrac{q^2}{b^2} + \cfrac{3}{4}f_H\left(\cfrac{z}{L}\right)^\frac{6}{\nu}e^{-c_Bz^2}\cfrac{q^2_H}{b^3}\\
\cfrac{T_{y_2y_2}}{g_{y_2y_2}} &=& -\cfrac{\phi'^2}{4L^2b}z^2g - \cfrac{V}{2} + \cfrac{f}{4}\left(\cfrac{z}{L}\right)^\frac{4}{\nu}\cfrac{q^2}{b^2}+ \cfrac{3}{4}f_H\left(\cfrac{z}{L}\right)^\frac{6}{\nu}e^{-c_Bz^2}\cfrac{q^2_H}{b^3}\\
\cfrac{T_{y_3y_3}}{g_{y_3y_3}} &=& -\cfrac{\phi'^2}{4L^2b}z^2g - \cfrac{V}{2} + \cfrac{f}{4}\left(\cfrac{z}{L}\right)^\frac{4}{\nu}\cfrac{q^2}{b^2}+ \cfrac{3}{4}f_H\left(\cfrac{z}{L}\right)^\frac{6}{\nu}e^{-c_Bz^2}\cfrac{q^2_H}{b^3}
\eea



\end{document}